\documentclass[%
 aip,
 pof,
 amsmath, amssymb,
 reprint,%
 floatfix,
]{revtex4-1}

\usepackage{graphicx}
\usepackage{dcolumn}
\usepackage{bm}
\usepackage{subfigure}

\usepackage{psfrag}
\usepackage{amsmath}
\usepackage{array}
\usepackage[]{color}

\DeclareMathSizes{7}{7}{5}{4} 
\DeclareMathSizes{5}{5}{3}{2} 

\definecolor{correction}{rgb}{1,0,0}
\definecolor{green}{rgb}{0,1,0}
\definecolor{comment}{rgb}{0,0,1}
\definecolor{refkey}{rgb}{1,0,0}
\definecolor{labelkey}{rgb}{1,0,0}
\definecolor{goetz}{rgb}{.7,.5,0}
\definecolor{francesco}{rgb}{.5,.5,0}

\newcommand{\lpar}{\left(}
\newcommand{\rpar}{\right)}
\newcommand{\lbk}{\left \lbrack}
\newcommand{\rbk}{\right \rbrack}

\newcommand{\ga}{{\alpha}}
\newcommand{\gb}{{\beta}}
\newcommand{\gc}{{\gamma}}

\DeclareMathOperator{\Tr}{Tr}

\newcommand{\beq}[1]{\begin{equation}\label{#1}}
\newcommand{\eeq}{\end{equation}}
\newcommand{\bal}[1]{\begin{align}\label{#1}}
\newcommand{\eal}{\end{align}}
\newcommand{\eref}[1]{Eq.~\eqref{#1}}
\newcommand{\figref}[1]{Fig.~\ref{#1}}

\newcommand{\vecc}[1]{\mathbf{#1}}

\begin{document}

\preprint{AIP/123-QED}

\title[Cavitation inception of a van der Waals fluid at a sack-wall obstacle]{Cavitation inception of a van der Waals fluid at a sack-wall obstacle}

\author{G. K\"{a}hler}

\author{F. Bonelli}

\author{G. Gonnella}

\affiliation{ 
Dipartimento di Fisica, Universit\`a di Bari, and INFN, Sezione di Bari, Via Amendola 173, 70126 Bari, Italy 
}%

\author{A. Lamura}
\affiliation{%
Istituto Applicazioni Calcolo, CNR, Via Amendola 122/D, 70126 Bari, Italy
}%

\date{\today}

\begin{abstract}
Cavitation in a liquid moving past a constraint is numerically
investigated by means of a free-energy lattice Boltzmann simulation
based on the van der Waals equation of state. The fluid is streamed past an obstacle
and, depending on the pressure drop between inlet and outlet, vapor formation underneath the corner of the sack-wall is
observed. The circumstances of cavitation formation are investigated and it is found that the local bulk pressure and mean stress are insufficient to explain the phenomenon. Results obtained in this study strongly suggest that the viscous stress, interfacial contributions to the local pressure, and the Laplace pressure are relevant to the opening of a vapor cavity. This can be described by a generalization of Joseph's criterion that includes these contributions.
A macroscopic investigation measuring mass flow rate behavior and discharge coefficient was also performed. As theoretically predicted, mass flow rate increases linearly with the square root of the pressure drop. However, when cavitation occurs, the mass flow growth rate is reduced and eventually it collapses into a choked flow state. In the cavitating regime, as theoretically predicted and experimentally verified, the discharge coefficient grows with the Nurick cavitation number.

\end{abstract}

\maketitle

\section{INTRODUCTION}

The analysis of cavitation is a significant concern for both fundamental fluid-dynamics and engineering applications such as high-pressure atomizers, spray generators, pumps, propellers, hydraulic turbines, etc. Cavitation is defined as the rupture of a liquid due to a pressure drop, which falls  below a certain critical value, at approximately constant liquid temperature, with the subsequent formation of vapor bubbles ~\cite{brennen-1995, dabiri-2007}. These bubbles collapse suddenly when they encounter a region with higher pressure ~\cite{koivula-2000}, thus releasing a large amount of energy in the form of shock and possibly light waves \cite{koivula-2000,brennen-1995,lohse-2005}. This phenomenon causes noise, vibrations and above all it can significantly affect the performance and damage the solid structures (cavitation erosion ~\cite{koivula-2000}) of engineering devices.
Despite the passing of more than one century since the first studies of 
Reynolds~\cite{reynolds-1873} and Parsons~\cite{parsons-1906} on the effects of cavitation on ship propellers, a comprehensive understanding of the physics of the phenomenon is still lacking as its analysis is challenging from both an experimental and a numerical point of view.

From an experimental point of view, cavitation is a relevant subject addressed by several researchers for both external \cite{kawanami-1997,weiland-2012,huang-2013} and internal flows \cite{winklhofer-2001,payri-2013,suo-2007,suo-2006,mishra-2005,tamaki-1998, hiroyasu-2000, payri-2004, de-giorgi-2013, gavaises-2009, giannadakis-2008}. 
In the field of high-pressure injection processes an inclusive study of this phenomenon is fundamental, since the disturbance induced by cavity bubbles promotes the jet atomization \cite{tamaki-1998,hiroyasu-2000}, thus affecting the mixing and combustion process.
In this context a very accurate and complete documentation of the flow phenomena, which occur in a transparent quasi two-dimensional real size nozzle, was provided by Winklhofer {\it et al.} \cite{winklhofer-2001}. 
In their work, the flow properties are described with both the conventional hydraulic characterisation, i.e., with data for mass flow as a function of the pressure drop, and with measurements of the vapor field, of the pressure field, and of the velocity profiles. The results show as the onset of cavitation is located near the wall at the entrance of the hole, specifically, in the shear layer between the incoming fluid, which separates from the entrance corner, and the recirculating fluid, which is near the nozzle wall. Velocity peaks at the boundary of the recirculation zone were found. These peaks are very large near the liquid-gas interface for choked flow conditions.
Payri {\it et al.} \cite{payri-2013} also studied the behavior of cavitation in a transparent real size cylindrical nozzle by using four different fluids. They performed both hydraulic characterization and flow visualization and showed that cavitation begins before the mass flow collapse. Additionally their experiments suggest that fluids with smaller viscosity tend to cavitate sooner.
Flow visualization and laser Doppler velocity measurements were carried out by Suo {\it et al.} \cite{suo-2007} at various conditions of cavitation and Reynolds numbers.
They assert that both the cavitation regime, inside the nozzle, and the liquid jet, near the nozzle exit, are strongly affected by the cavitation number and less by the Reynolds number. 
The role of surface tension effects on cavitation in microdevices was pointed out by 
 Mishra and Peles \cite{mishra-2005}. They examined cavitating flows through a two-dimensional micro-orifice, $11.5 \mu m$ wide, and found that with respect  to macroscale counterparts a much larger effort is needed to promote cavitation.

From a numerical point of view, cavitation is part of the wider family of multiphase flows~\cite{kuo-acharya-2012}. In this context, several efforts have been addressed in order to develop models able to capture the complex nature of such flows \cite{chung-2004,niu-2006,darbandi-2010,giannadakis-2008,yeom-2006}.  The main difficulties, dealing with multiphase flows,  are related to the presence of an interface between the different phases, which quickly changes its shape and where huge variations of the fluid properties are located. 
Traditional approaches are the Eulerian-Eulerian and Eulerian-Lagrangian modeling, with the vapor phase treated as a continuum or a discrete phase, respectively. The Eulerian-Eulerian approach requires  `jump conditions', in order to take into account the sharp changes across the interface, and techniques, such as the volume of fluid (VOF) \cite{hirt-1981}, to track the interface.
 Another widely used approach for the simulation of cavitating flows, is the homogeneous equilibrium flow model \cite{wang-2001, utturkar-2005,tseng-2010}. This model considers the two phases perfectly mixed, thus solving the conservation equations for the whole mixture. The liquid (or vapor) mass (or volume) fraction is introduced to model the density change \cite{utturkar-2005}. The problem can be closed by using some specific equation that couples density with pressure
 or a transport equation for the liquid or gas phase
~\cite{wang-2001, utturkar-2005, tseng-2010}. The latter approach requires the definition of source and sink terms which are modelled by using either empirical correlation, interfacial dynamics or bubble dynamics \cite{tseng-2010}.

In the last decades several studies \cite{chung-2004, niu-2006, darbandi-2010, giannadakis-2008, salvador-2013, goncalves-2014} made use of these models in order to simulate cavitating flows.
Niu and Lin \cite{niu-2006} developed a multiscale model in order to predict slurry erosion in ducts subjected to cavitating flows. In their work an Eulerian two-phase mixture model is used to simulate the cavitating flow field. The model employs three equations of state for water, vapor, and saturation states in order to take into account the phase-change process.
Darbandi and Sadeghi \cite{darbandi-2010} examined the fluid dynamic behavior of a cavitating flow through a sharp-edged circular orifice, by using a homogeneous approach,  in order to untangle the question whether it is necessary to consider the effect of non-condensable gas when dealing with cavitation.
Another important study,  based on the Eulerian-Lagrangian approach, was performed by Giannadakis {\it et al.} \cite{giannadakis-2008}. Here, cavitation is triggered by pre-existing nuclei and a full Rayleigh-Plesset equation is used to simulate the bubble growth and collapse dynamics. The authors claim that for the first time a model which takes into account the intrinsic stochastic nature of cavitation and also physical phenomena such as bubble breakup, coalescence and turbulent dispersion, is presented. The predictive capability was assessed through comparisons with experimental data obtained in real-size and enlarged models of diesel nozzle holes.
For the sake of brevity and considering that the body of literature in this field is quite large, we prefer to refer the reader to other publications (Refs. ~\onlinecite{brennen-1995, knapp-1970, lecoffre-1999, wang-2001, utturkar-2005, tseng-2010} ), where an extensive review of many aspects of cavitation can be found from both physical and numerical point of view.

The above mentioned studies have clarified many features of cavitation processes from the phenomenological point of view. However,  fundamental aspects of cavitation, such as the physics that underlies cavitation inception, remain not satisfactorily understood. Important questions concern the role played by  viscous stress  in the formation of the first cavitating bubbles.
Indeed, while in the literature it is typically assumed that the bulk value of the pressure determines the creation of a cavitating bubble, Joseph \cite{joseph-1998}  argued that, in fact, the significant parameter is the  total stress where both bulk pressure and viscous stress are taken into account. 
Experiments are not able to analyse this phenomenon, since, in most of the experimental tests, nucleation has a stochastic behavior due to its heterogeneous nature. Numerical simulations therefore appear to be a more promising approach for the purpose of comparing the traditional pressure criterion with Joseph's expression~\cite{joseph-1998}. In this context, it would be also important
to clarify the role of interfacial energies, contributing to the total stress balance, and expected to be relevant in any process of bubble formation\cite{bray-1994,onuki-2002}.
	
We explore in this work the potential of the lattice Boltzmann method (LBM)\cite{succi-2001, succi-2015,benzi-1992, gonnellayeomans-2009} 
in simulating cavitating flows. 
We analyze  the flow and thermodynamic behavior underlying the inception of  cavitation in a constrained geometry
and we will examine the validity of Joseph's criterion.
LBM is an algorithm that solves the Boltzmann equation in a discretized phase space. 
The method allows to recover the continuity and the Navier-Stokes equations in the continuum limit at the
second order in the Knudsen number. This method has been widely applied in several contexts to study simple and complex
fluids \cite{duenweg-2009}. Here the main benefit of the LBM approach is that it does not need an {\it ad hoc} model to capture the distinctive features of cavitating flows, such as the phase transition process, with its multiscale nature, and the dynamically varying liquid-vapor interface. Indeed, by using an appropriate non-ideal equation of state \cite{sofonea-2004, cristea-2006, gonnella-2007, gonnella-2009, cristea-2010}, which takes into account the real fluid behavior, cavitation emerges spontaneously from the dynamical approach \cite{falcucci-2013}.
In our simulations we considered homogeneous cavitation induced by a fast flow past a sack-wall obstacle. This standard geometry is widely used in numerical studies of cavitation \cite{dabiri-2007,darbandi-2010,giannadakis-2008, falcucci-2013, falcucci-2013a}. It can be considered a benchmark problem in order to have a comprehensive understanding of flow-induced cavitation as well as a simplified geometry useful for analyzing high pressure injectors.

LBM was first used for studying cavitation by Sukop and Or \cite{sukop-2005}. In that paper the cavitation of a single bubble was induced on a square geometry by decreasing the  external pressure on top and bottom walls
in absence of any imposed flow. A similar study was later addressed in Refs.~\onlinecite{xpchen-2011,zhong-2012}.
Only recently, the much more complex problem of cavitation in constrained flow environments has been successfully 
investigated \cite{falcucci-2013a,falcucci-2013}. Falcucci {\it et al.} \cite{falcucci-2013} applied a lattice 
Boltzmann model with a two-belt implementation \cite{falcucci-2007} of a Shan-Chen pseudo-potential \cite{shan-1993} to a  geometry similar to the one considered in this work and found cavitation underneath the obstacle. 
They relate their results to the stress criterion put forward by 
Joseph \cite{joseph-1998} once the cavitation threshold is properly
modified by considering surface-tension contribution.
In the model of Falcucci {\it et al.} \cite{falcucci-2013} 
the chosen pseudo-potential mimics a non-ideal equation of state.
Here we use  a lattice Boltzmann equation (LBE) model \cite{coclite-2014} 
which is derived by using a
Gauss-Hermite projection of the continuum equation \cite{shan-2006}.
A body force in the LBE \cite{guo-2002, tiribocchi-2009, gonnella-2010, gonnella-2011}, 
based on the van der Waals (vdW) free-energy functional\cite{rowlinson-1982},
guarantees that the fluid locally satisfies the van der Waals equation 
of state. 
The present approach allows to rigorously obtain 
in the continuum limit the continuity and the generalized 
Navier-Stokes equations without spurious terms. However, the presence of unphysical spurious velocities at
interfaces cannot be avoided and a general 9-point second-order finite difference scheme to compute the spatial derivatives was used.
This approach is known to  reduce these unwanted effects~\cite{shan-2006b,sbragaglia-2007}.
Another advantage of this model is that the phase diagram does not depend
on the relaxation time of the LBE which controls fluid viscosity.
Moreover, the surface tension is properly included in the pressure tensor 
so that it would be  interesting  
to analyse the inception of cavitation and the role
of stress and interfacial contributions by our method.

Additionally, for the first time, we employ LBM for direct hydraulic characterization of the constrained duct and compare the results with those obtained in experimental studies in similar geometries.

The present work is organized as follows: In section \ref{sec:MODEL} the mathematical model is introduced, together with some remarks on the implementation. In section \ref{sec:CAVITATION CRITERIA} a local and a global criterion for predicting and measuring cavitation are introduced. Numerical results for the former are discussed in section \ref{sec:CAVITATION INCEPTION} and for the latter in section \ref{sec:MACROSCOPICCAVITATION}. Finally, the conclusions are summarized.

\section{Model and Implementation}
\label{sec:MODEL}

We briefly describe  here the  two-dimensional isothermal 
lattice Boltzmann model for non-ideal fluids introduced  in Ref.~\onlinecite{coclite-2014}.
More details on the method and on the algorithm implementation are given in Ref.~\onlinecite{coclite-2014} and in Appendix A.
The ability of the model in reproducing the correct phase diagram of a van der Waals fluid and the surface tension behavior are also described in Ref.~\onlinecite{coclite-2014}.

\subsection{Lattice Boltzmann Equation for a van der Waals Fluid}
The time evolution of the system is defined by means of the dimensionless
lattice Boltzmann equation (LBE)
\begin{eqnarray}
f_i(\vecc{x}+\vecc{v}_i \Delta t, t+\Delta t) - f_i(\vecc{x}, t) = 
\frac{\Delta t}{\tau} \left\lbrack f_i^0(\vecc{x},t) - f_i (\vecc{x}, t) 
\right \rbrack + \Delta t F_i, \;\;\;\;\;\; i=0,...,8
\label{eqn:LBE1}
\end{eqnarray}
where the distribution functions $\{f_i\}$ 
are associated to the lattice velocities $\{\vecc{v}_i\}$ at lattice 
site $\vecc{x}$ and discrete time $t$. 
Here the square lattice $D2Q9$ is considered, 
where 9 distribution functions are defined
at each node with links to nearest and next-to-nearest 
neighbors.
The forcing terms $\{F_i\}$ depend on the
the force density $\vecc{\mathcal{F}}$ acting on the fluid and 
have to be conveniently determined,
$\tau$ is the relaxation time, and $\Delta t$ is the time step.
The equilibrium distribution functions $\{f_i^0\}$ are given by a 
second-order Hermite expansion of the Maxwell-Boltzmann distribution function
\cite{shan-2006}
\begin{eqnarray}
f_i^0(\rho, \vecc{u}, \theta) = \rho w_i 
\Big \{ 1 + v_{i,\ga} u_\ga 
+ \frac{1}{2}\lbk u_\ga u_\gb (v_{i, \ga} v_{i, \gb} - \delta_{\ga\gb}) 
+ (\theta - 1)(v_i^2-2) \rbk \Big \} \label{eqn:f0},
\end{eqnarray}
where $\rho = \sum_i f_i $ is the fluid density, $\rho \vecc{u} = 
\sum_i f_i  \vecc{v}_i + \frac{\Delta t}{2} \vecc{\mathcal{F}}$ is the fluid 
momentum, and $\theta$ is the temperature, which is not a dynamical variable in this model.
 The Hermite expansion fixes the values $|\vecc{v}_i|= \Delta x / \Delta t = \sqrt{3}$ with $\Delta x$ being the lattice unit (lu) for horizontal and vertical links with $i=1,...,4$,
$|\vecc{v}_i|=\sqrt{6}$ for diagonal links with $i=5,...,8$, $|\vecc{v}_0|=0$ 
for the rest velocity, and the weights $w_i=1/9$ for $i=1,...,4$, $w_i=1/36$ 
for $i=5,...,8$, and $w_0=4/9$.

The forcing term $F_i$ in \eref{eqn:LBE1} is given by \cite{guo-2002,coclite-2014}
\begin{eqnarray}
&&F_i =  w_i \lpar 1-\frac{\Delta t}{2\tau}\rpar \Big \{ 
v_{i,\ga} \mathcal{F}_\ga +  
\frac{1}{2} [ u_\ga \mathcal{F}_\gb + u_\gb \mathcal{F}_\ga + \lpar 1 - \theta \rpar 
\nonumber \\
&& \times \lpar u_\ga \partial_\gb \rho + u_\gb \partial_\ga \rho + \partial_\gc 
\lpar \rho u_\gc \rpar \delta_{\ga\gb} \rpar ]  \lpar v_{i,\ga} v_{i,\gb} - 
\delta_{\ga\gb}\rpar \Big \}.
\label{eqn:forcing}
\end{eqnarray}

In order to describe a van der Waals fluid the force has to be
\beq{eqn:force}
\mathcal{F}_\ga = \partial_\ga \lpar p^i - p^w \rpar 
+ \kappa \rho \partial_\ga \nabla^2 \rho  = \partial_\ga p^i 
- \partial_\gb \Pi_{\ga\gb},
\eeq
where $p^i=\rho \theta$ is the ideal gas equation of state (EOS), 
\beq{eqn:vdwEOS}
p^w=\frac{3\rho\theta}{3-\rho} - \frac{9}{8}\rho^2
\eeq
is the van der Waals EOS with
critical point at $\rho_c=\theta_c=1$, and
\beq{eqn:piab}
\Pi_{\ga\gb} = \lbk p^w - \kappa \rho \nabla^2 \rho 
- \frac{\kappa}{2}\lpar \nabla \rho \rpar^2 \rbk \delta_{\ga\gb} 
+ \kappa \partial_\ga \rho \partial_{\gb} \rho
\eeq
is the pressure tensor.
It can be computed from the free-energy
functional \cite{rowlinson-1982}
\begin{equation}
\Psi=\int d{\bf r} \Big [ \psi(\rho,\theta) +
\frac{\kappa}{2} (\nabla \rho)^2    \Big ]
\end{equation}
with bulk free-energy density given by
\begin{equation}
\psi=\rho \theta \ln \Big (\frac{3 \rho }{3 -\rho}\Big )-\frac{9}{8}\rho^2.
\end{equation}
The term proportional to  $\kappa$ takes into account the energy cost for
the formation of interfaces and allows to change the surface tension
independently of the temperature.

Chapman-Enskog expansion at second order shows that the continuity equation
\beq{eqn:cont}
\partial_t \rho + \partial_\ga (\rho u_\ga) + O(\partial^3) = 0
\eeq
and the Navier-Stokes equation
\begin{eqnarray}
\partial_t (\rho u_\ga) + \partial_{\gb} \lpar \rho u_{\ga} u_{\gb} \rpar = 
- \partial_\gb \Pi_{\ga\gb}
+ \partial_{\gb} \lbk \eta \lpar \partial_{\ga} u_{\gb} + 
\partial_{\gb} u_{\ga} \rpar \rbk + O(\partial^3) \label{eqn:ns}
\end{eqnarray}
can be properly recovered with 
$\eta = \rho \lpar\tau-\frac{\Delta t}{2}\rpar$ being the fluid viscosity. 

\subsection{Geometry and Boundary Conditions}

The model is used to simulate a fluid flowing in a channel from the left 
to the right past a sack-wall obstacle. This geometry is frequently adopted 
in numerical simulations of cavitation \cite{darbandi-2010, falcucci-2013a, falcucci-2013}. 

The simulated system consists of a rectangular simulation box where 
the ratio of length $L_x$ and height $L_y$ is given by $L_x/L_y = 3/2$. 
The fluid moves into the system from the left hand side
and, after travelling for a distance $L_x/3$, 
finds an obstacle of height $L_y/2$ which is connected to the top 
of the channel. 
The channel is thus constrained to a height $h=L_y/2$, half of the initial one, for the 
remaining length $(2/3) L_x$.

In the following we will use two different sets of inflow/outflow boundary conditions (BC).
We will refer to the first choice as fixed-density BC and 
to the second one as fixed-pressure BC. 
In the first case a fixed density $\rho_{\text{us}}$ and a fixed 
velocity ${\bf u}_{\text{us}}$ are used at the inflow layer. On this layer 
the distribution functions 
are given by the values of the equilibrium distribution
functions
$f^0_i(\rho_{\text{us}}, \vecc{u}_{\text{us}}, \theta)$. 
At the outflow a zero-gradient velocity condition is enforced with fixed 
density $\rho_{\text{ds}}$. In general we use $\rho_{\text{us}}= \rho_{\text{ds}}= \rho_0$ 
where $\rho_0$ is the initial density condition applied to all of the bulk fluid domain.

To compare with experimental conditions \cite{payri-2013, nurick-1976}, 
also fixed-pressure BC at inlet and outlet are considered. 
At the inflow the total upstream pressure $p_{\text{us}}$, 
which is the sum of the kinetic contribution and of the van der Waals pressure, 
is fixed and the $x$-component of the velocity is calculated enforcing
a zero-gradient velocity condition.
At the outflow a similar approach is used but the fixed-pressure used at the boundary $p_{\text{ds}}$ 
is just the van der Waals one. (More details on the implementation of the inlet and outlet boundary conditions
are given in Appendix A.1) 

At walls and corners
no-slip BC with neutral wetting and local density conservation 
are enforced \cite{lamura-2001,tiribocchi-2011} (see details in Appendix A.2).

\subsection{Thermodynamic Consistency of the Model}

In all simulation runs the lattice unit $\Delta x$ was fixed to 1 and 
the value $\Delta t=\sqrt{3}/3$ was used unless specified otherwise. 
Several quantities, which will be later useful, were measured
at different values of the temperature $\theta$ and of the
interface term $\kappa$ and we will give their values in the following.
They will be expressed in units of $\Delta x$, $\Delta t$, $\rho_c$, and $\theta_c$.

In table I the equilibrium vapor $\rho_{V,sim}$ and liquid densities $\rho_{L,sim}$ measured in simulations are compared to the corresponding analytic values $\rho_{V,an}$ and $\rho_{L,an}$, respectively.
Runs were performed on a lattice of size $100\times100$ where a vapor droplet of radius $25$ at density $\rho_{V,an}$ was immersed in liquid at density $\rho_{L,an}$ and let to equilibrate.
Both equilibrium vapor and liquid densities are closer to the analytic values at larger $\theta$ and $\kappa$.

\begin{table}
\begin{tabular}{c|c|c|c|c|c}
$\theta$ & $\kappa$ & $\rho_{L,an}$ & $\rho_{L,sim}$ & $\rho_{V,an}$ & $\rho_{V,sim}$\\
\hline
0.9 & 0.0 & 1.65727 & 1.59266 & 0.425742 & 0.39654  \\
    & 0.1 &         & 1.60999 &          & 0.40368  \\
    & 0.2 &         & 1.62043 &          & 0.40830  \\
    & 0.3 &         & 1.62731 &          & 0.41143  \\
\hline
0.85& 0.0 & 1.80714 & 1.7067  & 0.31973  & 0.28552  \\
    & 0.1 &         & 1.7331  &          & 0.29401  \\
    & 0.2 &         & 1.7492  &          & 0.29967  \\
    & 0.3 &         & 1.7599  &          & 0.30353  \\
\hline
0.8 & 0.0 & 1.93270 & 1.81931 & 0.23966  & 0.20405  \\
    & 0.1 &         & 1.82595 &          & 0.21424  \\
    & 0.2 &         & 1.85206 &          & 0.22049  \\
    & 0.3 &         & 1.86695 &          & 0.22476  
\end{tabular}
\caption{Values of the equilibrium vapor and liquid densities, $\rho_{V,sim}$ and $\rho_{L,sim}$, measured from simulations, 
are compared to the analytical values $\rho_{V,an}$ and $\rho_{L,an}$ 
for different values of the temperature $\theta$ and of the interface term $\kappa$.
}
\label{tab:eosdata}
\end{table}

Table II reports the maximum density $\rho_{\text{spin}}$ from $1D$ and $2D$ simulations at which spinodal decomposition 
on the liquid branch was observed.
For this purpose a very slightly perturbed system is initialized at a 
density close to the analytic spinodal value and it is observed whether 
the system phase separates. $\rho_{\text{spin}, 1D}$ was measured in a 
$100\times1$ lattice while $\rho_{\text{spin}, 2D}$ in a $50\times50$ system.
The analytic value of the liquid branch spinodal density 
is well reproduced in both the one- and 
the two-dimensional cases. In Appendix B we investigated the impact of shear on the spinodal density. The results suggest that shear does not significantly impact the following results on cavitation inception of section \ref{sec:CAVITATION INCEPTION}.

\begin{table}
\begin{tabular}{c|c|c|c|c}
$\theta$ & $\kappa$ & $\rho_{\text{spin},an}$ & $\rho_{\text{spin},1D}$ & $\rho_{\text{spin},2D} $\\
\hline 
0.9 & 0.0 & 1.39160 & 1.3911 & 1.3895 \\
    & 0.1 &         & 1.3907 & 1.3881 \\
    & 0.2 &         & 1.3904 & 1.3868 \\
    & 0.3 &         & 1.3901 & 1.3856 \\
\hline
0.85& 0.0 & 1.48880 & 1.4884 & 1.4873 \\
    & 0.1 &         & 1.4881 & 1.4862 \\
    & 0.2 &         & 1.4879 & 1.4852 \\
    & 0.3 &         & 1.4876 & 1.4842 \\
\hline
0.8 & 0.0 & 1.57428 & 1.5739 & 1.5731 \\
    & 0.1 &         & 1.5737 & 1.5722 \\
    & 0.2 &         & 1.5735 & 1.5713 \\
    & 0.3 &         & 1.5733 & 1.5705 
\end{tabular}
\caption{Values of liquid-branch spinodal density from $1D$ 
($\rho_{\text{spin},1D}$) and $2D$ ($\rho_{\text{spin},2D}$)
simulations are compared 
to the analytic values $\rho_{\text{spin},an}$ 
for different values of the temperature $\theta$ and of the
interface term $\kappa$.
}
\label{tab:spindata}
\end{table}

Surface tension was measured directly through the free energy in 
a one-dimensional system. For this purpose two quantities were computed:
\begin{itemize}
\item The bulk free energy 
\beq{eqn:psibulk}
\Psi_{\text{bulk}} = \int d{\bf r} \psi(\rho, \theta) 
\eeq
calculated with half of the system in the pure liquid phase and with the other half in the pure vapor phase. The densities used are the numerical equilibrium values reported in table I.
\item The total free energy including gradient contributions after the system 
had fully~{equilibrated}
\beq{eqn:psitot}
\Psi = \int d{\bf r} \Big [ \psi(\rho, \theta) 
+ \frac{\kappa}{2} \lpar \nabla \rho \rpar^2 \Big ] .
\eeq
\end{itemize}
The difference $\Psi-\Psi_{\text{bulk}}$
would then yield the total free energy change due to the interface
including numerical contributions
that cannot be attributed to the $\kappa$-term, but that effectively act as an interface energy contribution.
The surface tension $\sigma_{\text{sim}}$ is then calculated as 
\beq{eqn:sigmasim}
\sigma_{\text{sim}} = \Psi - \Psi_{\text{bulk}}.
\eeq

Additionally the surface tension was also calculated in a 1D system according to 
its definition
\beq{eqn:sigmakappa}
\sigma = 
\int dx \frac{\kappa}{2} \lpar \nabla \rho \rpar^2
\eeq
by using the numerical values of $\rho$ in the equilibrated system 
($\sigma_{\kappa,sim}$)
and also the approximate equilibrium profile \cite{wagner-2007} 
($\sigma_{\kappa,\text{num}}$)
\beq{eqn:psitanh} 
\rho(x) = \rho_{V,sim} + \frac{\rho_{L,sim} - \rho_{V,sim}}{2}\lbk 1 
+ \text{tanh}
\lpar x \sqrt{\frac{\frac{1}{\theta}-1}{2 \kappa} } \rpar \rbk.
\eeq
The results are given in table III. 
For all three temperatures, the values of $\sigma_{\kappa,\text{sim}}$ 
and $\sigma_{\kappa,\text{num}}$ coincide fairly well and the agreement 
improves with growing $\kappa$. 
The numerical contribution $\sigma_{\text{sim}}(\kappa=0)$ 
added to any value of $\sigma_{\kappa,\text{sim}}$ or 
$\sigma_{\kappa,\text{num}}$ for non-zero $\kappa$ is expected to give the equivalent 
surface tension $\sigma_{\text{sim}}$ for non-zero $\kappa$. 
This appears to be reasonably consistent, moreso for higher temperature $\theta$ and lower surface tension $\kappa$.

\begin{table}
\begin{tabular}{c|c|c|c|c}
$\theta$ & $\kappa$ & $\sigma_{\text{sim}}$ & $\sigma_{\kappa,\text{sim}}$ & $\sigma_{\kappa,\text{num}}$  \\
\hline
0.9 & 0.0 & 0.0418  & 0       & 0      \\
    & 0.1 & 0.0634  & 0.0361  & 0.0272 \\
    & 0.2 & 0.0814  & 0.0516  & 0.0493 \\
    & 0.3 & 0.0971  & 0.0636  & 0.0681 \\
\hline
0.85& 0.0 & 0.0738  & 0       & 0      \\
    & 0.1 & 0.1124  & 0.0649  & 0.0472 \\
    & 0.2 & 0.1449  & 0.0929  & 0.0868 \\
    & 0.3 & 0.1733  & 0.1150  & 0.1208 \\
\hline
0.8 & 0.0 & 0.1555  & 0       & 0      \\
    & 0.1 & 0.1665  & 0.0968  & 0.0687 \\
    & 0.2 & 0.2157  & 0.1403  & 0.1277 \\
    & 0.3 & 0.2589  & 0.1739  & 0.1791  
\end{tabular}
\caption{Measured values of the surface tension $\sigma$ 
in a one-dimensional system for different values of the temperature 
$\theta$ and of the
interface term $\kappa$ (see the text for details).}
\label{tab:sigma}
\end{table}

\section{Cavitation Prediction Criteria}
\label{sec:CAVITATION CRITERIA}

Cavitation criteria are quantitative relations used to predict the occurrence of cavitation.
Macroscopic criteria are based on  the hydraulic characterization of the flow  and on the definition of the cavitation number $C_n$ 
(see below). They are  frequently used in engineering applications.
Microscopic criteria are based on the behavior of local variables. They  relate the local stress to a maximum pressure at which the cavity can be formed.
We describe now two cavitation criteria and some relations  which were investigated numerically and that will be useful
in the following.

Hydraulic characterization originates in the experimental field. This is in part because when dealing with engineering devices, direct local detection of cavitation is difficult due to lack of visual access to the relevant area of the experiment \cite{koivula-2000}. Then cavitation can be  detected by looking at the stationary mass flow rate as a function of the pressure drop. Indeed, for an incompressible flow, the theoretical mass flow rate obtained from Bernoulli's equation is directly proportional to $\left(p_{\text{us}}-p_{\text{ds}}\right)^{1/2}$. However, the actual mass flow rate is smaller than the theoretical one because of losses due to boundary layer, vena contracta, turbulence and cavitation. Therefore, 
it is common to define a discharge coefficient as the ratio between the actual flow rate and the theoretical one 
\begin{equation}
C_d = \frac{\dot{m}}{h \left[ 2 \rho_L \left(p_{\text{us}}-p_{\text{ds}}\right) \right]^{1/2}},
\end{equation}
where $\dot{m}$ is the actual mass flow rate and $\rho_L$ the equilibrium liquid density. 
Experimental works \cite{winklhofer-2001,payri-2013} showed that by fixing $p_{\text{us}}$ and decreasing $p_{\text{ds}}$ mass flow rate grows with $\left(p_{\text{us}}-p_{\text{ds}}\right)^{1/2}$ until it reaches the so-called critical cavitation point after which, though decreasing $p_{\text{ds}}$, mass flow remains constant and the flow is called choked. Therefore, 
by looking at the behavior of the discharge coefficient,
it is possible to distinguish a region where cavitation is absent or negligible and another region where cavitation significantly reduces or completely removes a further increase in mass flow rate. 

When the flow is not cavitating, $C_d$ is primarily a function of the Reynolds number, which is defined as
\begin{equation}
Re=\frac{u_{th} h}{ \nu},
\end{equation}
where $u_{th} = \sqrt{2(p_{\text{us}} - p_{\text{ds}})/\rho_L}$ is the theoretical Bernoulli velocity measured at the end of the channel and $\nu=\eta/\rho_L$ is the kinematic viscosity. In contrast, for cavitating flow, the cavitation number we will define is the main influence on $C_d$ \cite{payri-2013,nurick-1976}.

The definition of the cavitation number is based on Bernoulli's equation. It relates the difference between the pressure at a certain location and the maximum pressure at which a cavity can be formed, often the vapor pressure $p_V$, to a dynamic pressure. However, the specific location at which the relevant observables are measured, and in particular, the geometry are not uniquely defined. Different approaches have been proposed \cite{brennen-1995, nurick-1976, payri-2013}. Following Nurick \cite{nurick-1976} we choose here a definition appropriate for a constrained orifice where the upstream pressure is considered and the dynamic pressure is expressed as $p_{\text{us}} - p_{\text{ds}}$:
\beq{eqn:CN2}
C_{n} = \frac{p_{\text{us}} - p_V}{p_{\text{us}} - p_{\text{ds}}}.
\eeq
This definition originates from the one-dimensional model discussed in Ref.~\onlinecite{nurick-1976} and allows the calculation of $C_d$ for a cavitating flow as
\begin{equation}
C_d =C_{c}\cdot \left(C_{n}\right)^{1/2}.
\label{eqn:nurick}
\end{equation}
Here $C_{c}$ is the contraction coefficient given by the ratio $h_c/h$, where $h_c$ is the vena contracta flow section.
The above is valid only in the presence of cavitation and therefore represents an extremely useful tool for indirect identification of cavitation in practical applications as also shown in the experimental work of Payri et al \cite{payri-2013}.

In contrast to the macroscopic considerations one can also establish local criteria based on the behavior 
of the total stress of the fluid, with respect to  some critical pressure, at a given position where 
cavitation might occur.
Joseph {\it et al.} \cite{dabiri-2007} argue that it is
 the largest directional stress overcoming a pressure threshold that gives rise to cavity formation.
For a simple fluid the total or Cauchy stress tensor  can be expressed by the following constitutive equation
\beq{eqn:totalstress}
\vecc{T} =  - p \vecc{1} + \boldsymbol{\tau},
\eeq
where $p$ is the bulk pressure, $\boldsymbol{\tau}$ is the viscous stress tensor, and $\vecc{1}$ is the unit tensor.
The original version of Joseph's maximum tension criterion states that the quantity to be 
 compared to a pressure threshold $p_c$
is  the largest eigenvalue $T_{11}$ of $\vecc{T}$. 
Cavitation would occur at points where
\beq{eqn:maxtension}
T_{11} + p_c > 0.
\eeq

Taking into account that $T_{11}$ is a negative number one may interpret this as follows: If the magnitude of the maximum principal stress drops below the critical pressure threshold $p_c$ for vapor formation, the formation of a cavity is expected.
This generalizes the usual assumption \cite{brennen-1995} that it is the pressure or the local mean stress 
that determines the formation of a cavity in a fluid.
The mean stress is defined in terms of the trace of $\vecc{T}$ as 
\begin{equation}
\label{eq:meanstress}
T^0 = \frac{1}{D} \Tr \vecc{T},
\end{equation}
where $D$ is the spatial dimension. For a Newtonian and incompressible fluid $\boldsymbol \tau$ is the stress deviator of the total stress tensor, since $\Tr \boldsymbol{\tau}~=~0$, and therefore the mean stress is equal to $-p$.

In the microscopic view there is one caveat here. The bulk pressure is not sufficient to express all pressure contributions, which also include interfacial terms.
For this reason the full pressure tensor \eref{eqn:piab} must be considered, 
so that we introduce as the total stress the expression
\beq{eqn:totalstressveccp}
\vecc{T}_{\vecc{\Pi}} = - \vecc{\Pi} + \boldsymbol{\tau},
\eeq
with the corresponding mean stress denoted as 
$T^0_{\vecc{\Pi}} = \frac{1}{D} \Tr \vecc{T}_{\vecc{\Pi}}$.
In this context it is natural to assume that also the pressure threshold 
is affected by interfacial contributions.
Falcucci {\it et al.} \cite{falcucci-2013} include  the 
 Laplace pressure  in the pressure threshold so that
\beq{eqn:pc}
p_c \left(R\right) = {\hat p} - \frac{\sigma}{R}.
\eeq
Here $\sigma$ is the surface tension and $R$ the minimal radius 
for the formation of a vapor bubble. 
 In experimental studies and numerical engineering works ${\hat p}$ 
is usually identified with the vapor pressure $p_V$ of the fluid. While this is a reasonable assumption in experimental investigations due to the almost inevitable presence of nucleation kernels,
 in this particular study the equilibrium vapor pressure is likely not a very good estimate. In the case of a fluid without any kind of nucleation kernels or other sources of local symmetry breaking such as the system used here, one expects ${\hat p}$ to be close to the spinodal pressure rather than the equilibrium vapor value.

\section{Cavitation Inception}
\label{sec:CAVITATION INCEPTION}
\subsection{Overview}

\begin{figure*}
\includegraphics[clip=true, width=.95\textwidth, viewport= 200 380 550 740]{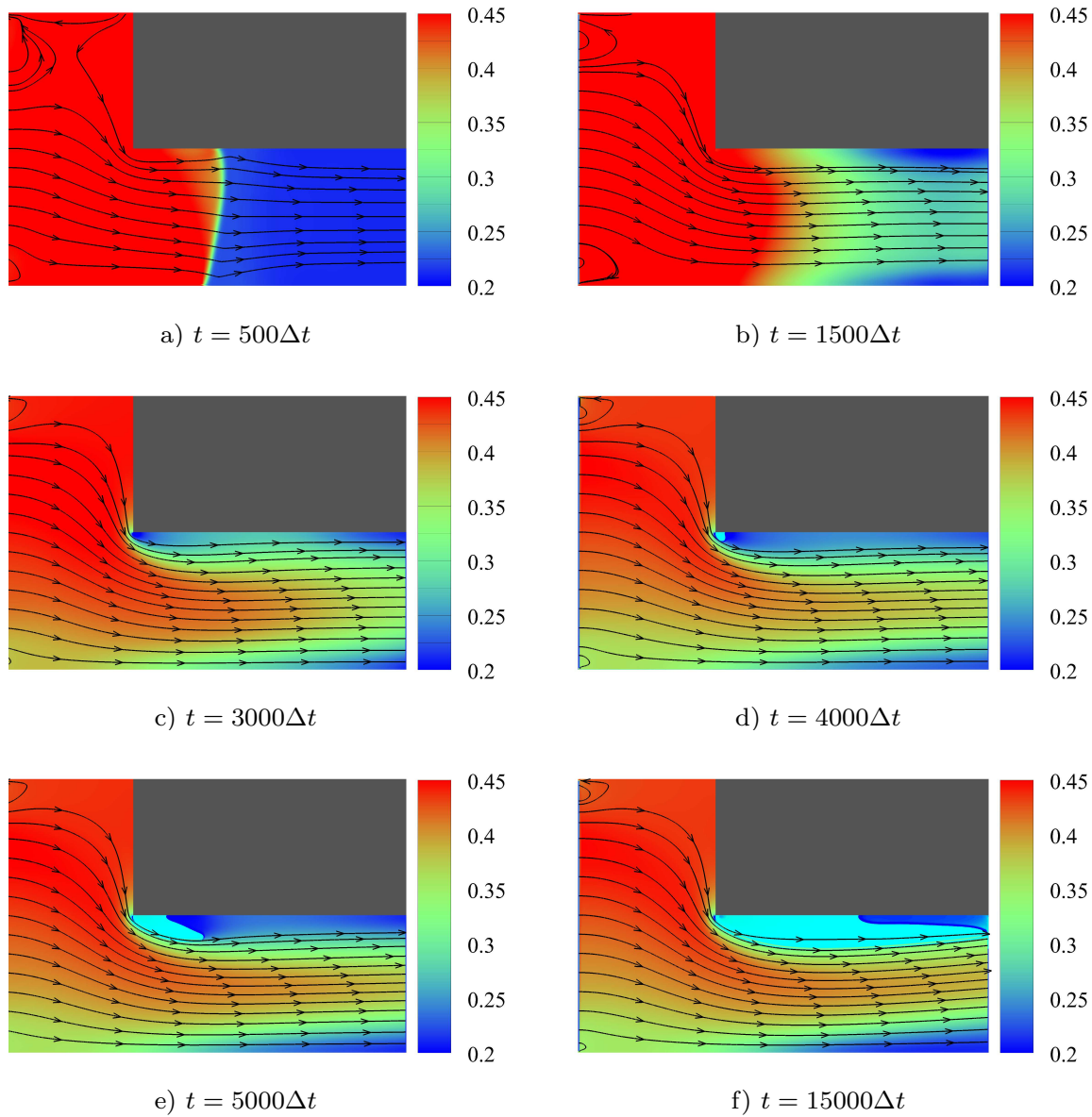}
\caption{Contour plots of the total pressure $p_{\text{tot}} = p^w + \frac{1}{2} \rho u^2$ at different times in simulation on  
a $601\times401$ lattice with fixed-density BC together with the flow field. The parameters($\theta=0.9, \rho_0 = 1.63, u_{\text{us}} = 0.25$) are chosen such that the typical phases of the cavitation can be observed. In a) the transient initial wave front propagating to the right is visible. b) shows the formation of the stable flow profile after the wave has made contact with the outflow boundary layer. In c) the flow profile has matured and the pressure depletion underneath the obstacle corner becomes visible in dark blue color. In d) the initial cavity has formed which grows in e) to finally, after a long time, extend to the end of the channel while covering a significant section of the lower obstacle boundary in f). The vapor fraction in d), e), and f) has been colored in cyan. The pressure of the vapor area is about $p_{V,\text{bubble}}=0.215$ and is between the equilibrium pressures calculated from the numerical equilibrium densities of table \ref{tab:eosdata} where $p_L = 0.202$ and $p_V = 0.234$. 
}
\label{fig:overview}
\end{figure*}
\begin{figure}
\includegraphics[width=.6\textwidth, angle=270]{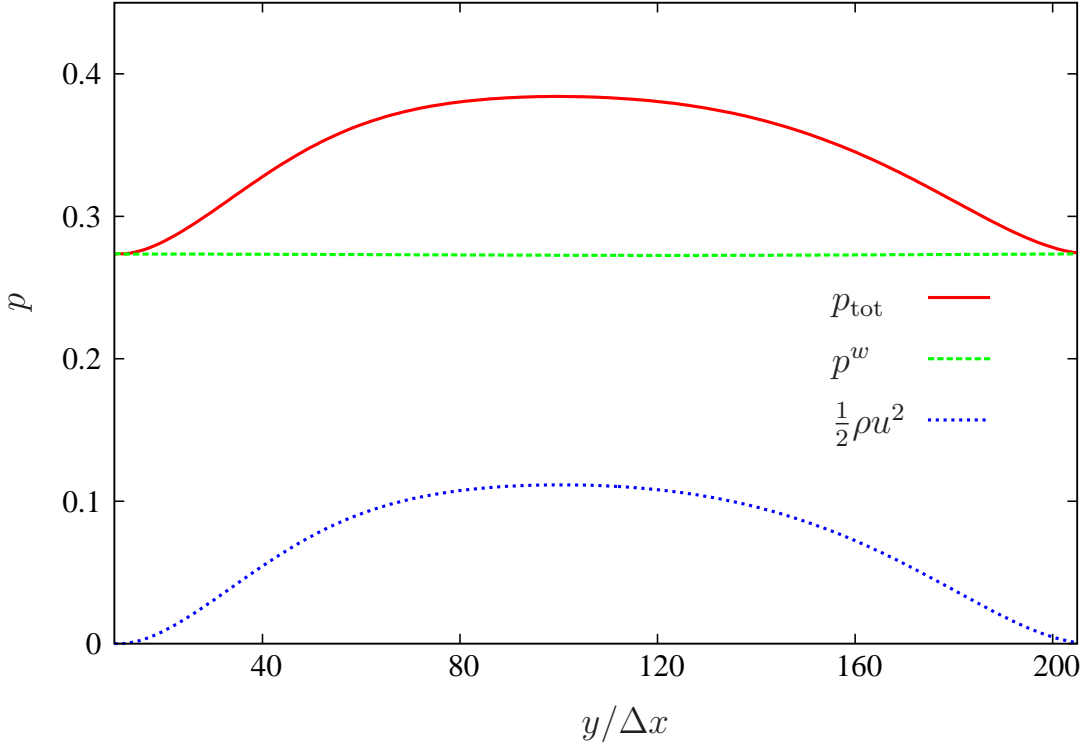}
\caption{Total, static and dynamic pressures as function of $y$ lattice position at $x=400$, i.e. half the constrained channel length, in the channel of 
\figref{fig:overview} at $t=2500 \Delta t$. Here, $y=0$ is the lower boundary of the channel whereas $y=200$ is at the wall of the obstacle.
}
\label{fig:pmix02500}
\end{figure}

\begin{figure}
\includegraphics[viewport = 190 515 570 600 ,clip=true, width=.95\textwidth]{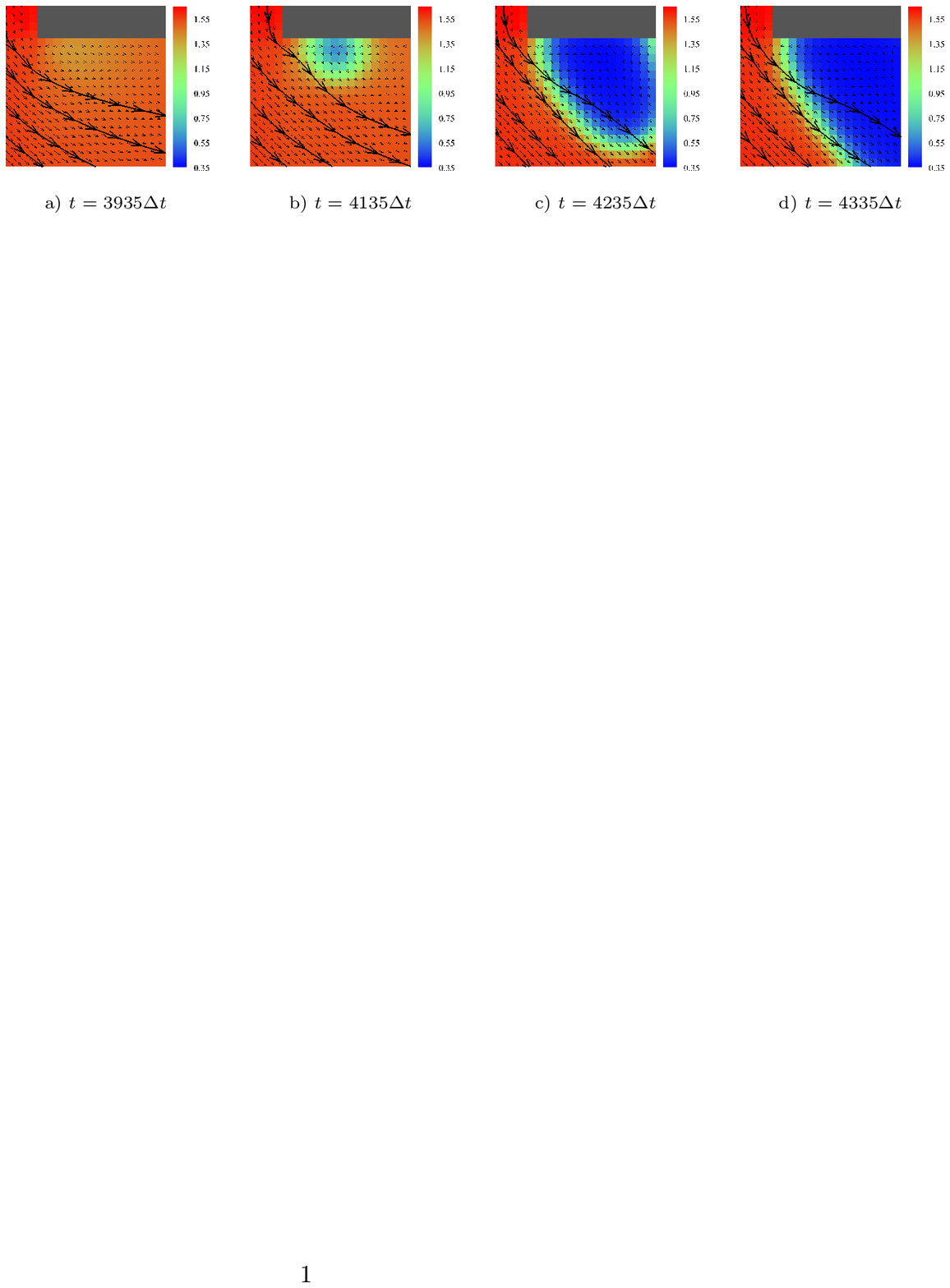}
  \caption{Contour plot of the density $\rho$ and macroscopic velocity ${u}$ shown at four different times shortly after the pressure reached the liquid-branch spinodal value. Simulation parameters: $1201\times 801$ mesh  with lattice spacing of $0.5 \text{lu}$ and time unit adjusted accordingly, fixed-density BC,  $\theta=0.9$, and inflow velocity $u_{\text{us}} = 0.15$.}
\label{fig:cavi2}
\end{figure}
As previously mentioned, the simulations are performed in a two-dimensional channel constrained by a sack-wall. The simulation box has the
size of $601\times401$ lattice units ($\text{lu}$). 
Resolutions of $1.5$ times and twice the number of lattice nodes with correspondingly reduced $\Delta x$ and $\Delta t$ have been tested and, where stability allows, the results are consistent with the findings at the standard resolution shown here. 

An example of the behavior of the system from initialization at rest to the formation of a stable velocity profile and eventually the formation of a cavity is given in \figref{fig:overview}. 
The total pressure $p_{\text{tot}} = p^w + \frac{1}{2} \rho u^2$ and the flow lines are given at different times. In this example, as well as in most measurements in this work, the temperature is $\theta=0.9$, the surface tension parameter $\kappa = 0.1$, and the viscosity $\nu = 0.71$. The inflow and outflow boundaries are computed according to the fixed-density BC at an upstream velocity of $u_{\text{us},x} = 0.25$ and $\rho_0 = 1.63$, which is close to the equilibrium liquid density. The static pressure corresponding to this density is $p^w = 0.2234$. Non-boundary nodes are initialized at rest with the same density $\rho_0$.
The fluid begins to stream in from the left hand side. The initial pressure front first collides with the obstacle wall and is, as seen in \figref{fig:overview} (a), partially reflected. The attenuation of the inflow velocity near the upper and lower channel walls, according to \eref{eqn:uxin}, reduces the pressure near the corners. Instabilities due to the rebounding pressure wave that are otherwise prominent are thus eliminated.

Once the pressure wave approaches the outflow boundary the zero velocity gradient boundary at the outflow layer allows material to leave the system to the right. As the channel walls enforce zero velocity the fluid near the center of the channel accelerates and the static pressure is reduced. This reduction first occurs near the outflow boundary and then propagates towards the left as the material in the channel begins to move. The outflow BC fix the density at the outflow layer $\rho_{\text{ds}} = \rho_0$ and, as $\rho_0 > \rho_V$, no vapor can originate from the outflow layer in this example. The stationary flow pattern of later stages of the simulation begins to form, see \figref{fig:overview} (b). 

The configuration after another 1500 time steps is shown in \figref{fig:overview} (c). Pressure and the flow pattern have become almost stationary and, as expected, the fluid velocity profile reaches a maximum at the center of the constrained channel. The static pressure is constant perpendicular to the flow direction and any change in total pressure is due to the increased dynamic pressure contribution near the center of the channel. \figref{fig:pmix02500} shows these pressure profiles in the $y$-direction at half the constrained channel length. At this time it is also visible that the lowest static pressure in the system is just underneath the obstacle. The pressure distribution and flow pattern observed here are rather typical for the sack-wall geometry (see, e.g., Ref.~\onlinecite{falcucci-2013}).

With the static pressure falling below a threshold value, a nucleation kernel is formed which then evolves into a vapor cavity. This happens first underneath the obstacle. This small initial cavity is first visible in \figref{fig:overview}~(d). In the last three images of \figref{fig:overview} the developed vapor is colored in cyan. The formation of this initial cavity is investigated in detail in section \ref{sec:INCEPTIONsub} and a higher resolution view of the density of the initial bubble formation is shown in \figref{fig:cavi2} albeit at lower inlet velocity than that of \figref{fig:overview}. 
In this example the area near the wall exhibits a pressure well below the equilibrium vapor pressure and thus the cavity immediately begins to grow. The exact shape and direction of this growth depend on the pressure or density distribution in the channel. The cavity typically grows towards the center of the channel first. Then, due to the increase of the velocity towards the center of the channel, it is stretched along the flow lines of the fluid as seen in \figref{fig:overview}~(e). The bubble continues to grow until it hits the outflow boundary layer. The de-wetting of the gas bubble at the obstacle is a slower process but in almost all observed cases, a sufficiently long running simulation will exhibit a fully formed gas bubble covering most of the lower obstacle surface. 
An intermediate state of this bubble growth is shown in \figref{fig:overview}~(f). In this example, at the point of smallest diameter for fluid flow, the vena contracta, less than $3/4$ of the channel width is available for fluid transport resulting in a reduced mass transport. This effect and its consequences are discussed in section \ref{sec:MACROSCOPICCAVITATION}.
We finally mention that, unlike in the computation results shown in \figref{fig:overview}, if the downstream pressure remains well above the equilibrium vapor pressure the de-wetting of the wall remains partial in our numerical experiments.

\subsection{Local Cavitation Inception}
\label{sec:INCEPTIONsub}

In order to understand the local mechanism of vapor formation and 
analyze the cavitation criteria discussed in section \ref{sec:CAVITATION CRITERIA}, the conditions under which the cavity forms were carefully investigated. 
\figref{fig:cavi2} mentioned in the previous section gives in double standard resolution the density and fluid velocity underneath the obstacle during initial bubble formation.
This higher resolution computation does not show significant quantitative or qualitative differences compared with the standard resolution case. 
In \figref{fig:cavi2} (a) the spinodal density is reached for the first time. The formation of the interface is 
initiated in \figref{fig:cavi2} (b) which then quickly progresses to the nucleation of a vapor cavity (\figref{fig:cavi2} (c) and (d)). It is worth noting that at this level of observation configurations that will not cavitate but have otherwise very similar parameters will be visually indistinguishable from that of \figref{fig:cavi2} (a). They will approach a state similar to \figref{fig:cavi2} (b) and will recede to a configuration similar to \figref{fig:cavi2} 
(a) where the density remains close to $\rho_{\text{spin}}$ but vapor formation does not occur.

We observe that prior to interface formation a single lattice site underneath the obstacle assumes the lowest density in the channel. 
This site is situated two lattice spacings below the obstacle and four lattice spacings to the right of the vertical wall of the obstacle for the resolution of $601\times401$ sites.
In the following, individual measurements of pressure and density with regards to cavitation inception are taken at this lattice position.
Cavitation is observed at all temperatures considered in table \ref{tab:eosdata}. Due to stability concerns, however, only cases of $\kappa \le 0.1$ are accessible at lower temperatures. For the best compromise between correct reproduction of the equilibrium phase diagram and the best stability it is chosen to limit the analysis of cavitation inception to $\theta=0.9$.
The macroscopic parameters available are the mean density $\rho_0$   
and the velocity imposed on the inflow boundary $u_{\text{us}}$.
For observing cavitation we choose
an initial density $\rho_0 = 1.55$ which is, as before, slightly
below the equilibrium liquid density 
and  in the metastable regime 
of the phase diagram, 
 and vary the inflow velocity. If not otherwise specified, the first set of BC with fixed-density at inlet and outlet is used.

\begin{figure}
\subfigure[\vspace{2mm}~$\kappa = 0.0$]{
\includegraphics[width = 0.33\textwidth, angle=270]{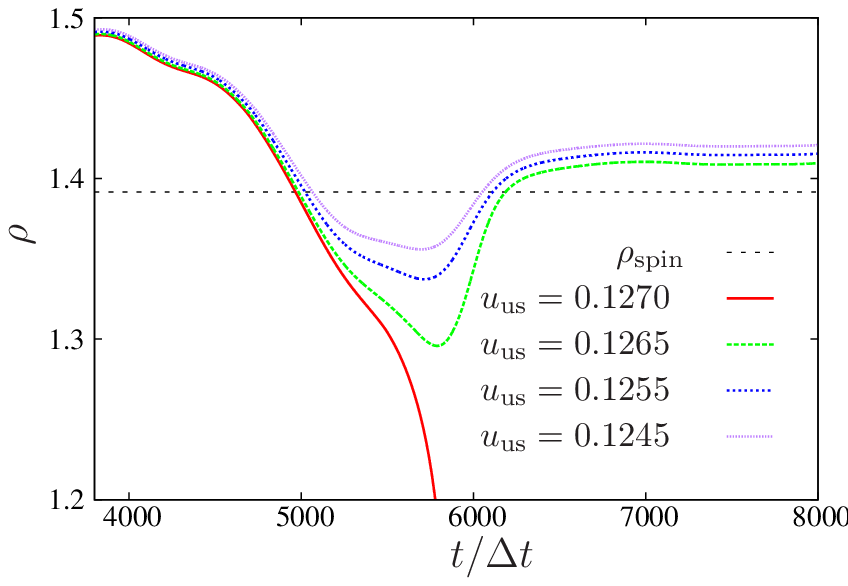}
}
\subfigure[\vspace{2mm}~$\kappa = 0.1$]{
\includegraphics[width = 0.33\textwidth, angle=270]{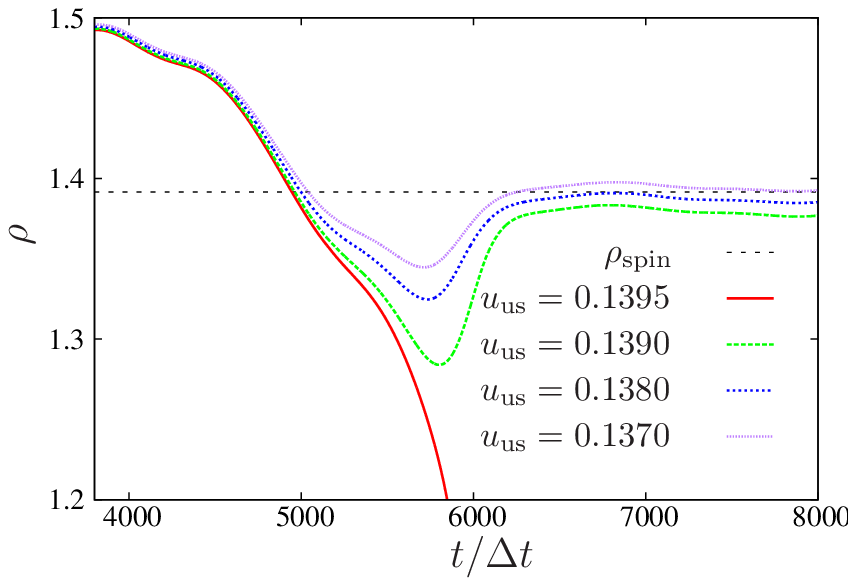}
}
\caption{ 
Density $\rho$ at the lowest density site during the ``transient cavitation measurement'' as a function of simulation time $t$ for different initial velocities $u_{\text{us}}$ for a) $\kappa=0$ and b) $\kappa = 0.1$. Cavitation only occurs for the highest $u_{\text{us}}$ shown (red continuous line). Other parameters: $\theta=0.9$, $\rho_0 = 1.55$.} 
\label{fig:rhotrans}
\end{figure}

\begin{figure}
\includegraphics[width = 0.5\textwidth, angle=270]{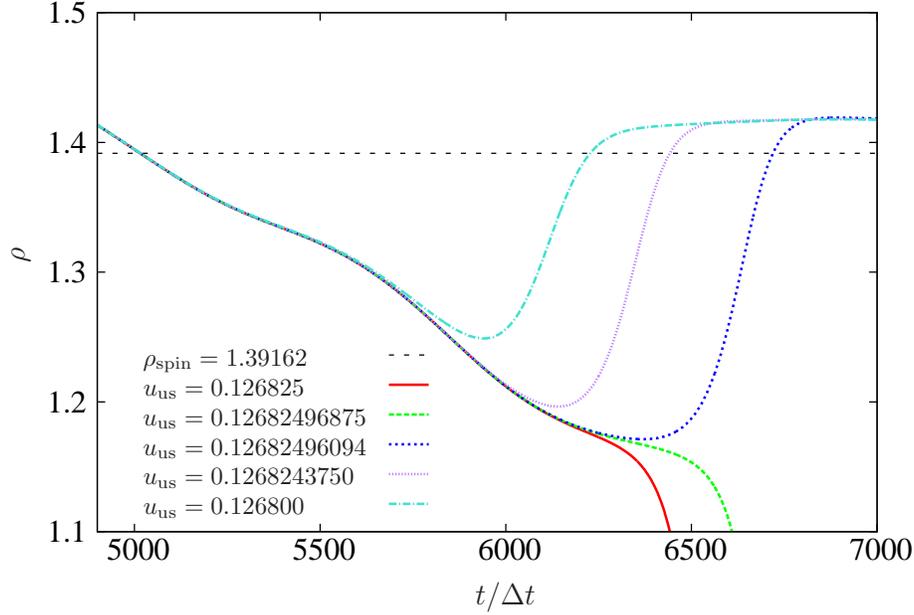} 
\caption{Density $\rho$ at the lowest density site during the ``transient cavitation measurement'' as a function of simulation time $t$ for different initial velocities $u_{\text{us}}$ at $\kappa=0$. Results here were obtained using an iterative approach minimizing the inflow velocity. 
A similar behaviour has been found at $\kappa=0.1$.
Other parameters: $\theta=0.9$, $\rho_0 = 1.55$.}
\label{fig:rhotransk00}
\end{figure}

In \figref{fig:rhotrans} the density at the lowest density site underneath the obstacle is given as a function of time for different inflow velocities.
In both cases, $\kappa = 0.0$ and $\kappa = 0.1$, 
 for sufficiently high inflow velocity, ($u_{\text{us}} \approx 0.11, 0.12 $ at $\kappa=0.0, 0.1$, respectively),
the density, independently of subsequent actual cavity formation, 
decreases below 
the spinodal value. As shown in \figref{fig:rhotrans}, if  $u_{\text{us}}$ is not large enough
this is a transient drop, 
with the density relaxing to a value close to the spindodal one at larger times.
A good estimate for the highest upstream velocity $u_{\text{us}}$ for which cavitation does not occur was found with  an iterative approximation.
 The minimal value of density obtained in this way is significantly below the liquid branch spinodal
 value of $\rho_{\text{spin}}(\theta=0.9) = 1.3916$ as seen in \figref{fig:rhotransk00} for the case with $\kappa = 0$.
It appears that cases can be constructed where the density temporarily drops very significantly below the liquid branch spinodal, but, due to the transient nature of this approach, no cavity is created. 
Likely the time for which this low density environment exists, is insufficient to allow for nucleation.
We call this protocol for observing cavitation ``transient cavitation measurement''.

\begin{figure}
\subfigure[~$\kappa = 0.0$]{
\includegraphics[width = 0.33\textwidth, angle=270]{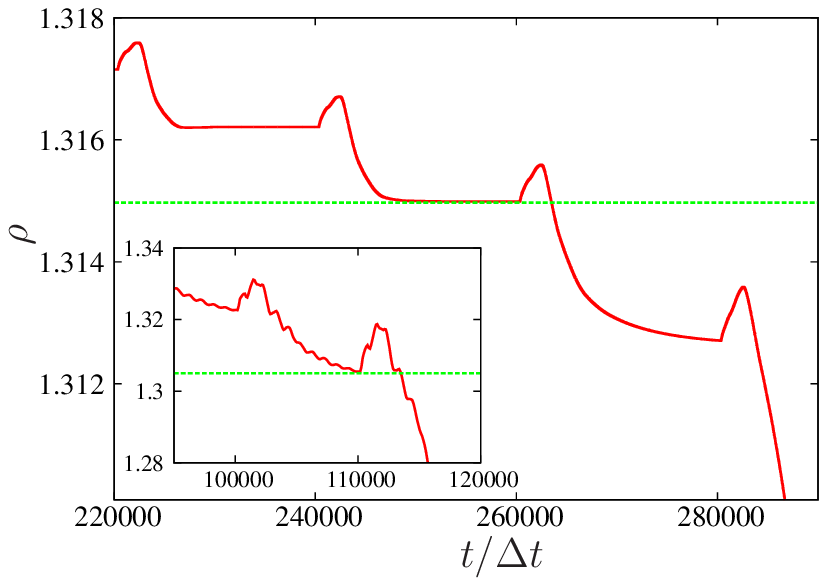}
}
\subfigure[~$\kappa = 0.1$]{
\includegraphics[width = 0.33\textwidth, angle=270]{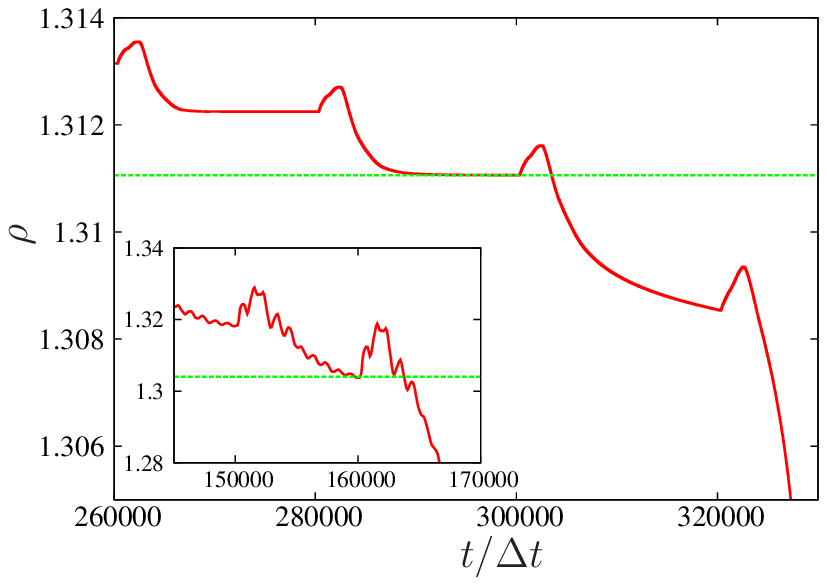}
}
\caption{Density $\rho$ measured at the lowest density site in simulation with fixed-density BC with velocity increments of $\Delta u_{\text{us}} = 0.0002$ every $2.0\times10^4$ time updates. The dashed lines indicate the lowest stable density observed and corresponds to the configuration shown in \figref{fig:minbubblek01}. Other parameters: $\theta = 0.9$, $\rho_0 = 1.55$. A similar measurement was conducted but with fixed-pressure BC although at much larger pressure steps. The inset plots do show that observed thresholds appear at similar values. Differences between inset and main figure are likely due to the much longer equilibration time of the fixed-pressure BC.}
\label{fig:rholong}
\end{figure}

To avoid the difficulties in quantifying the density threshold by the above approach,
the  protocol for observing  incipient cavitation was altered.
The system is initialized at an inflow velocity 
small enough that the system is  not cavitating in the initial transient approach.
Then, to establish a sequence of stationary states not affected by transient behavior,
the simulation is run for a time of order $10^4 \Delta t$, see \figref{fig:rholong}, significantly exceeding the one needed for transient cavitation formation,  and  the inflow velocity is then increased by a small $\Delta u_{\text{us}}$. This process, that we call "`steady-state cavitation measurement"', is repeated until cavity formation is observed. The lowest value for which the density still converges in the liquid regime is considered the closest pre-cavitation value we can attain and is used as a threshold. These threshold measurements are given in the main plot of \figref{fig:rholong} for fixed-density BC.

Similar measurements were performed using the alternative fixed-pressure inlet and outlet boundary conditions (discussed in Appendix A), varying the upstream pressure instead of the upstream velocity.
 These results are shown in the insets of \figref{fig:rholong}. 
For both types of boundary conditions we observed that the threshold densities are well below the liquid branch spinodal value.

The above results indicate that the simple assumption that cavitation occurs when the fluid density or pressure reach the liquid branch spinodal values is insufficient. Taking the liquid branch coexistence pressure or density as critical threshold would further increase the difference between the measured value and this critical value. 
One might  wonder if  the spinodal values listed in Table \ref{tab:spindata}, referring  to a system at rest, would be changed by flow.  It is known that multiphase fluids under applied flow, shear for example, exhibit a critical point depending on the flow rate \cite{gonnella-2000, gonnella-1998, onuki-2002}.
In our model, the jump in the horizontal velocity component underneath the obstacle could be roughly approximated by a linear shear profile (see \figref{v_and_rho_profiles} in the following).
Thus we decided to check how much the spinodal values of our system are affected by a shear flow. 
The  results of  these simulations are summarized in Appendix IB. We find  that the spinodal values measured at rest
 are changed very little by shear or, in any case, in a way that cannot explain the cavitation threshold measured above.

\begin{figure}
\includegraphics[width = 0.5\textwidth, angle=270]{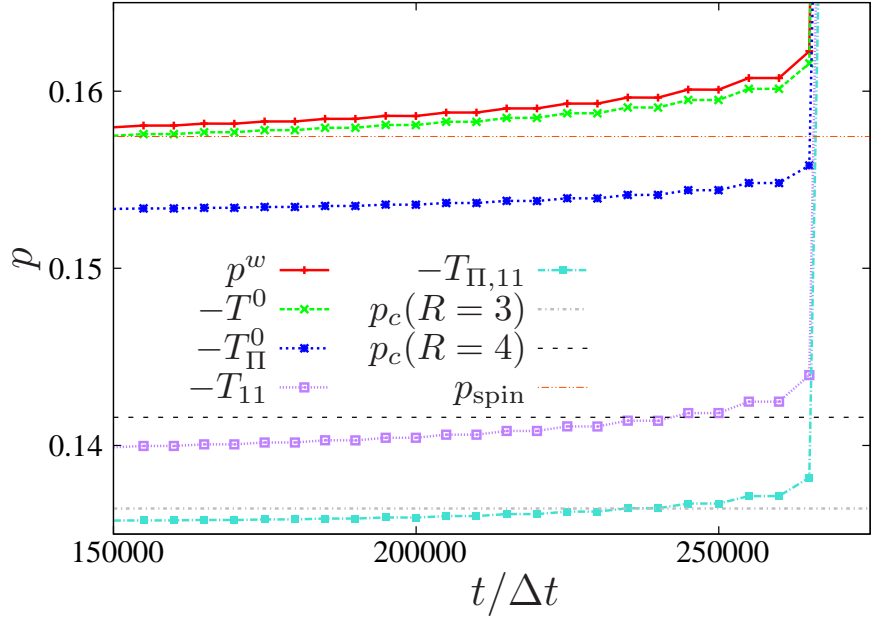}
\caption{Evaluation of the Joseph criterion measured in a $601\times401$ $\text{lu}$ simulation at the cavity inception site underneath the obstacle. Each data point corresponds to one stationary state of the type shown in \figref{fig:rholong} after increasing the upstream velocity. The last stable threshold is found at $t=260000 \Delta t$ which corresponds to an inflow velocity of $u_{\text{us}} = 0.144$. 
Other parameters: $\theta=0.9$, $\rho_0 = 1.55$, $\kappa = 0.1$.}
\label{fig:josephk01}
\end{figure}

One could also consider, instead of the pressure, the behavior of the  mean stress $T^0$ defined by (\ref{eq:meanstress})
and compare this with the spinodal critical pressure value. From \figref{fig:josephk01} one sees that the behavior of the mean stress
$T^0$ is very close to that of the static pressure $p^w$  so that also  this quantity cannot be useful 
for explaining the formation of cavitation. We observe  that the van der Waals pressure remains above the spinodal 
value as predicted by the analytical expression of van der Waals isotherms.

The Joseph criterion or its generalizations  discussed in Sect. \ref{sec:CAVITATION CRITERIA},
based on the largest stress eigenvalue and also including interface contributions,
 offer a natural way to  further analyze the formation of cavitation in terms of local variables.
 Before doing this, since interfacial contributions also affect the critical threshold \eqref{eqn:pc} through the Laplace term, we will  evaluate the typical size of the  low density region that, under proper conditions,  will evolve into a vapor  bubble.

\begin{figure}
\includegraphics[width = 0.4\textwidth, angle=0]{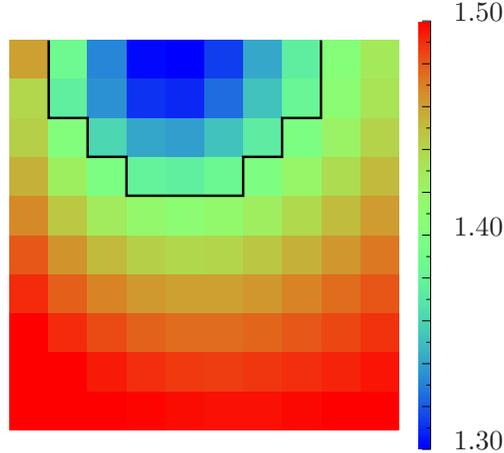}
\caption{Contour plot of the density observed directly underneath the obstacle at cavitation inception in a $601\times401$ $\text{lu}$ simulation at $\theta = 0.9$, $\kappa = 0.1$ and kinematic viscosity $\nu = 0.71$, i.e. $\tau = 1$. The black line indicates the threshold of the spinodal density $\rho_{\text{spin}}=1.39160$. }
\label{fig:minbubblek01}
\end{figure}

A magnified view of the density underneath the obstacle at the time of the threshold of \figref{fig:rholong} (b) is given in \figref{fig:minbubblek01}. This configuration is the lowest stable density prior to cavitation formation
and can be used to estimate the minimum droplet radius for evaluating the Laplace pressure contributions.
The  area inside the contour contains all lattice nodes for which $\rho<\rho_{\text{spin}}$. This half circular shape has a diameter of $D = 2R \approx 6lu$ so for this study it is considered $R \approx 3\text{lu}$.

The maximum tension criterion of Joseph of \eref{eqn:maxtension} compares the magnitude of the maximum stress with a critical pressure.
The maximum tension for both stress definitions (\eref{eqn:totalstress} and \eref{eqn:totalstressveccp}), and also the 
behavior of the mean stress $T^0_{\vecc{\Pi}}$
 are shown in \figref{fig:josephk01}.
Moreover, two different critical thresholds 
corresponding to different Laplace contributions in 
 \eref{eqn:pc} are plotted.
The Laplace pressure contribution is calculated by using the surface tension values $\sigma_{\text{sim}}$,
evaluated as described by~\eref{eqn:sigmasim}, and the critical radii
 $R=3$ and $R=4$.

According to the criterion one would expect cavitation when the magnitude of the local stress is smaller than the pressure threshold. In \figref{fig:josephk01} one can observe that the magnitude of the maximum stress eigenvalue, for both stress definitions, is significantly below the spinodal pressure also before cavitation occurs. Therefore the Joseph criterion with the critical threshold given by the spinodal pressure has not any particular meaning in our context.
On the other hand, the spinodal pressure adjusted by the Laplace pressure with a minimum droplet radius of $R \approx 3 \text{lu}$ appears in close vicinity to the maximum stress magnitude $T_{\Pi, 11}$.
So it appears that both the maximum stress eigenvalue and the critical pressure with the Laplace contribution  are relevant and in this particular experiment they are found in close vicinity to each other. Furthermore there is an appreciable difference between the two definitions of \eref{eqn:totalstress} and \eref{eqn:totalstressveccp} indicating that the surface tension contributions due to gradient terms in the pressure tensor should be considered relevant as well. 
\footnote {The final increase in the stress seen before cavitation is due 
to the contribution of the van der Waals pressure that,  following the van der Waals isotherm, increases when the fluid density decrease below the liquid spinodal value.}
We also remark  that the behavior of the mean stress $T^0_{\vecc{\Pi}}$ does not appear relevant in this context.

Simulations at higher (doubled) resolution confirm the above picture with the stress behaving as in \figref{fig:josephk01}.
For example, at $\kappa=0.01$, the magnitude of the maximum stress eigenvalue $T_{\Pi,11}$ almost reaches the pressure threshold $p_c(R=3)=0.1430$ before increasing due to cavitation with a plateau at $0.1448$ and $p_c(R=4)=0.1466$  (at $\Delta x=1$ the differences for the values of the pressure thresholds with respect to the case at higher resolution are negligible and the stress plateau is at $0.1445$).

Similar measurements were performed for the fixed-pressure inflow BC at three different viscosities: $\nu = 0.5$, $\nu = 0.71$, and $\nu = 0.9$ and confirm the above picture with a critical threshold diminishing with increased viscosity.

\section{Macroscopic Cavitation Prediction}
\label{sec:MACROSCOPICCAVITATION}

In this section cavitation is examined by using the standard approach of hydraulic characterization described in section \ref{sec:CAVITATION CRITERIA}. This technique is widely used in experimental works, such as by Payri {\it et al.} \cite{payri-2013} and by Winklhofer {\it et al.} \cite{winklhofer-2001}, in order to analyse the behavior of high pressure systems. Both, Payri {\it et al.} \cite{payri-2013}, and Winklhofer {\it et al.} \cite{winklhofer-2001}, used an axial-symmetric and a quasi two-dimensional nozzle, respectively, monitored the mass flow rate under steady flow conditions by keeping constant the upstream total pressure and varying the downstream static pressure. Thus they obtained data sets for mass flow rate as a function of the pressure drop independent of the total upstream conditions.
The results show that mass flow rate increases by increasing the pressure drop up to a maximum value after which even further decreasing the downstream pressure  mass flow remains constant. At this point the flow is said to be choked. Mass flow saturation is mainly caused by cavitation \cite{winklhofer-2001,payri-2013} and it accompanied by a reduction of the discharge coefficient \cite{payri-2013}. Specifically, Payri {\it et al.} \cite{payri-2013}, who performed a comprehensive analysis by considering four different fluids with various viscosities in order to cover a wide range in terms of Reynolds number, showed that under non-cavitating conditions mass flow rate grows asymptotically with the Reynolds number whereas under cavitating conditions, ~\eref{eqn:nurick} is met. Therefore, both mass flow choking and discharge coefficient reduction are indirect evidence of cavitation. Moreover, they showed that less viscous fluids tend to cavitate sooner. 
  
The contribution of this work is to test, for the first time, the applicability of the LBM for the analysis of the cavitation phenomenon by using a macroscopic approach closer to engineering practice which relies on observing the results in terms of mass flow rate versus pressure drop and discharge coefficient versus both cavitation and Reynolds number.

\begin{figure}[t]
\includegraphics[scale=0.4]{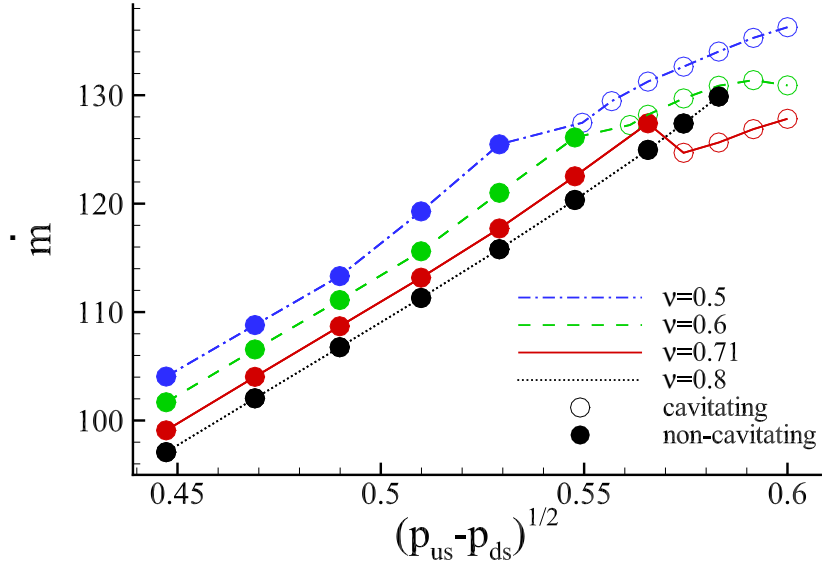}
\caption{Stationary mass flow rate $\dot{m}$ as a function of the square root of the pressure drop $\left(p_{\text{us}}-p_{\text{ds}}\right)^{1/2}$ for different values of the kinematic viscosity. Simulation parameters: lattice sites $601 \times 401$ $\text{lu}$ with fixed-pressure BC at $\theta=0.85$, $\kappa = 0.1$ and total inlet pressure $p_{\text{\text{us}}}=0.4$.}
\label{Mass_vs_deltap}
\end{figure}

Measurements reported in this section are obtained in simulations with the fixed-pressure BC discussed in Appendix \ref{app:bc}.
We used the base resolution of $601 \times 401$ $\text{lu}$, at $\theta=0.85$ and $\kappa = 0.1$, to perform several simulations by fixing the total inlet pressure
$p_{\text{us}}=0.4$, and varying $p_{\text{ds}}$ for
kinematic viscosities $\nu=0.5, 0.6, 0.71, 0.8$. 
The results in terms of stationary mass flow rate as a function of the square root of the pressure drop are given 
in Fig. ~\ref{Mass_vs_deltap}. At $\nu=0.5$ the mass flow rate increases almost linearly up to $p_{\text{ds}} = 0.12$ (last right filled circle on the dash-dotted line
in Fig. ~\ref{Mass_vs_deltap}). Subsequently, further reducing $p_{\text{ds}}$ the growth rate is reduced. A similar behavior is observed with kinematic viscosity equal to $0.6$. Here the mass flow rate increases almost linearly up to $p_{\text{ds}} = 0.10$ (last right filled circle on the dashed line in Fig. ~\ref{Mass_vs_deltap}), 
then the growth rate is reduced and, finally, the system collapses into a choked flow state. Mass flow choking is well visible for the case with kinematic viscosity equal to $0.71$ when the critical cavitation point is reached at $p_{\text{ds}}=0.08$ (last right filled circle on the continuous line 
in Fig.~\ref{Mass_vs_deltap}). Indeed, further reducing the downstream pressure the mass flow rate shows a small reduction
at $p_{\text{ds}} = 0.07$ and then settles to values comparable with those achieved at $p_{\text{ds}} = 0.08$.

%
\begin{figure}[t]
\subfigure[~$p_{\text{ds}}=0.08$] {\includegraphics[scale=.3,viewport= 50 50 640 400, clip]{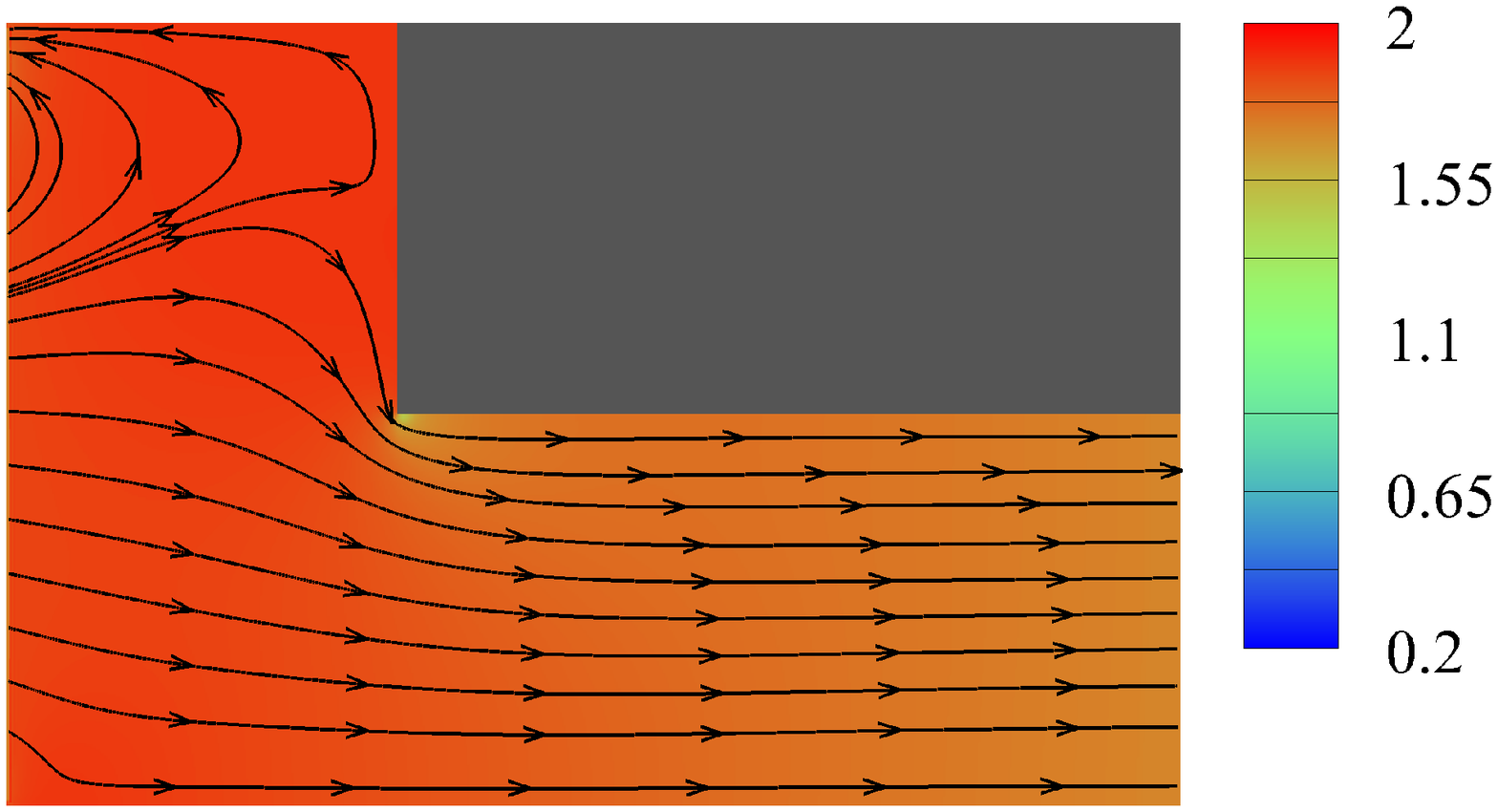}}
\subfigure[~$p_{\text{ds}}=0.07$]{\includegraphics[scale=0.3,viewport= 50 50 640 400, clip]{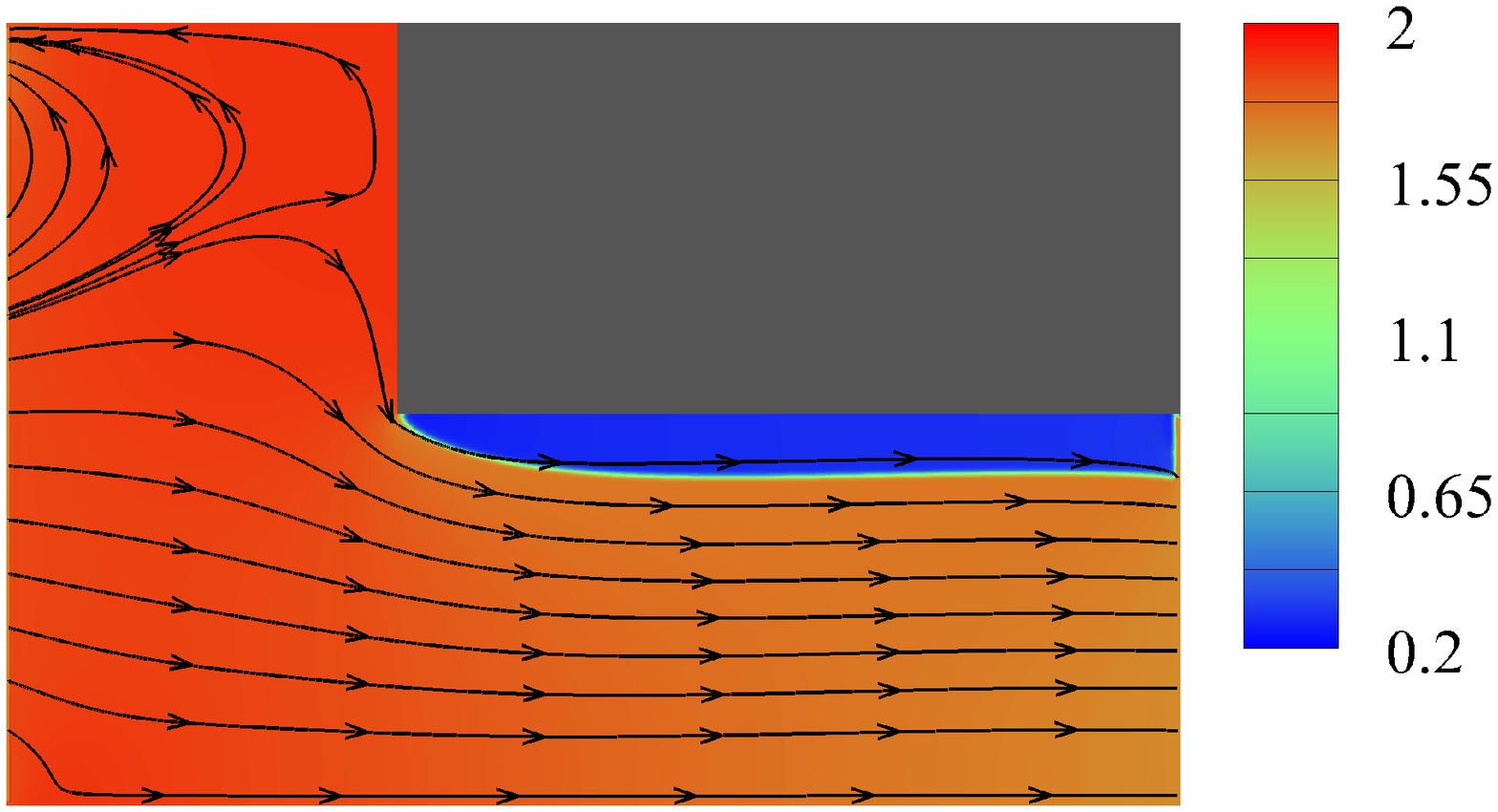}}
\caption{Density contour plots with flow lines. Simulation parameters: lattice of $601 \times 401$ $\text{lu}$ with fixed-pressure BC at $\theta=0.85$, $\kappa = 0.1$, $p_{\text{us}}=0.4$ and $p_{\text{ds}}=0.08$ (a), $p_{\text{ds}}=0.07$ (b) with kinematic viscosity equal to $0.71$.}
\label{density_contour_plot}
\end{figure}

%
The mass flow choking is clearly due to cavitation. This can be seen in~\figref{density_contour_plot}, where the density contour plots for the cases with $\nu=0.71$ and $p_{\text{ds}}=0.07, 0.08$ are given. 
There is no vapor with $p_{\text{ds}}=0.08$, whereas, when 
$p_{\text{ds}}=0.07$ vapor extends attached to the wall, along the entire length of the obstacle. A similar observation is also valid for the cases with kinematic viscosities equal to $0.5$ and $0.6$. Moreover, the flow lines, reported in the same figure, show that in both cases there is a significant recirculation zone near the front side of the obstacle.       
%

Finally, by considering again Fig. ~\ref{Mass_vs_deltap}, when the kinematic viscosity is increased to $0.8$, the mass flow rate always increases, almost  linearly, 
 with the square root of the pressure drop and it does not show any reduction of the growth rate up to the smallest $p_{\text{ds}}$ considered, which for this kinematic viscosity is $0.06$. A further reduction of $p_{\text{ds}}$ would
compromise the numerical stability. The absence of a reduction in the mass flow growth rate is due to the fact that up to $p_{\text{ds}}=0.06$ cavitation does not occur.

%
\begin{figure}[t]
\includegraphics[scale=0.4]{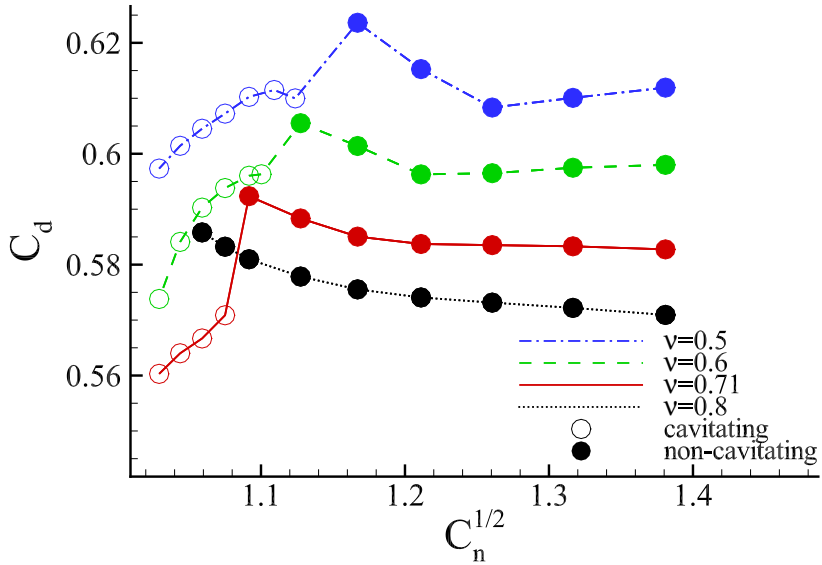}
\caption{Discharge coefficient $C_d$ as a function of the square root of the Nurick \cite{nurick-1976} cavitation number $C_{n}^{1/2}$ for different values of the kinematic viscosity. Simulation parameters: lattice sites $601 \times 401$ $\text{lu}$ with fixed-pressure BC at $\theta=0.85$, $\kappa = 0.1$ and total inlet pressure $p_{\text{\text{us}}}=0.4$.}
\label{Cd_vs_Cn2}
\end{figure}
%
%
Obviously, the reduction of the mass flow growth rate causes a decrease of $C_d$ as shown in Fig.~\ref{Cd_vs_Cn2}, where the discharge coefficient as a function of the square root of the Nurick
 cavitation number of \eref{eqn:CN2} is given (here $C_{n}$ is determined considering the liquid branch spinodal pressure instead of the vapor pressure in \eref{eqn:CN2}). Indeed, except for the case with kinematic viscosity equal to $0.8$, we can identify a critical cavitation number below which there is an abrupt reduction of $C_d$. In the non-cavitating regime, for cavitation numbers  higher than the critical value
(filled circles in Fig.~\ref{Cd_vs_Cn2}),  $C_{d}$ slightly increases by decreasing $C_{n}$ while it reaches
 a plateau at large $C_n$. 
In the cavitating regime,  moving towards the critical value of the cavitation number,
for kinematic viscosity equal to $0.5$ and $0.6$, $C_d$ shows  a small undershoot before  reaching 
the critical point. The undershoot is not present when the kinematic viscosity is $0.71$.
 These findings are qualitatively in good agreement with the experimental ones \cite{payri-2013} since it is possible to identify a critical point below which the discharge coefficient abruptly decreases. 
In this region $C_d$ is a  linear function of $C_{n}^{1/2}$, as suggested by
\eref{eqn:nurick} only for the case with $\nu=0.71$, whereas ~\eref{eqn:nurick} is not met when the kinematic viscosity is equal to $0.5$ and $0.6$. 

\begin{figure}[t]
\includegraphics[scale=0.4]{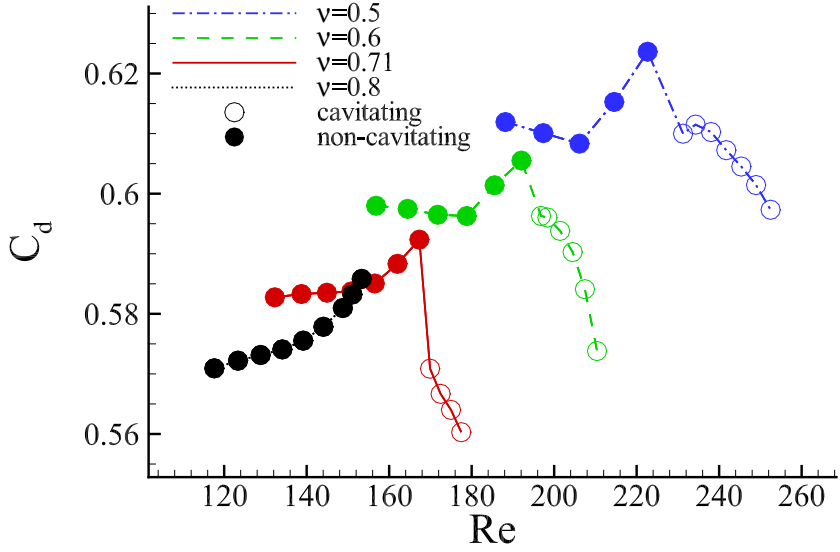}
\caption{Discharge coefficient $C_d$ as a function of the Reynolds number for different values of the kinematic viscosity. Simulation parameters: lattice of $601 \times 401$ $\text{lu}$ with fixed-pressure BC at $\theta=0.85$, $\kappa = 0.1$ and total inlet pressure $p_{\text{\text{us}}}=0.4$.}
\label{Cd_vs_Re}
\end{figure}

\begin{figure}[t]
\includegraphics[scale=0.4]{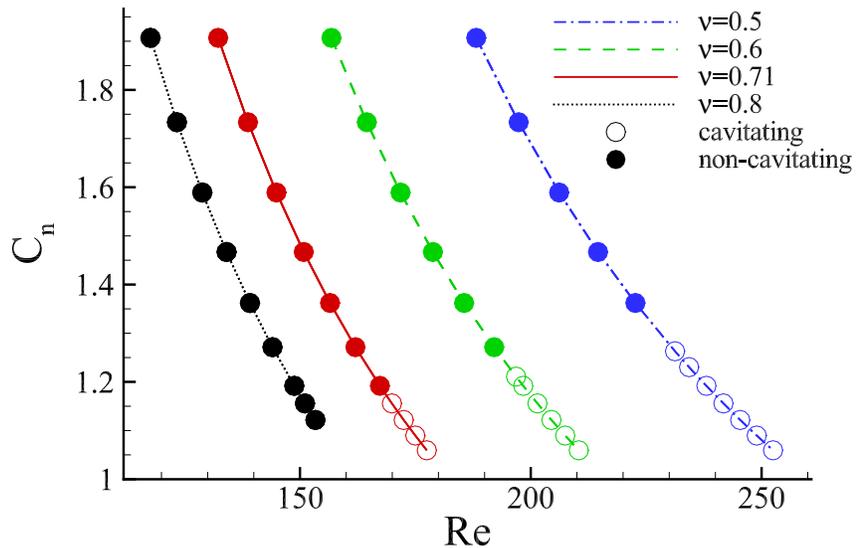}
\caption{Nurick cavitation number $C_{n}$ as a function of the Reynolds number for different values of the kinematic viscosity. Simulation parameters: lattice of $601 \times 401$ $\text{lu}$ with fixed-pressure BC at $\theta=0.85$, $\kappa = 0.1$ and total inlet pressure $p_{\text{\text{us}}}=0.4$.}
\label{Cn2_vs_Re}
\end{figure}

The partial disagreement with the Nurick law of \eref{eqn:nurick} can be due to the effects of the Reynolds number which are not taken into account  in the derivation of \eref{eqn:nurick}.
We show   the dependence of $C_d$ and $C_n$ on $Re$. 
In Fig.~\ref{Cd_vs_Re}, where the discharge coefficient is given as a function of $Re$, one sees that 
$C_d$ reaches a maximum and then it abruptly decreases due to cavitation effects. This result is qualitatively in good agreement with the experimental one by Payri et. al \cite{payri-2013}. Fig.~\ref{Cn2_vs_Re}
shows that $C_{n}$ decreases with $Re$ in both cavitating and non-cavitating cases.
 Given the dependence of  $C_d$ and $C_n$ on the Reynolds number,
it is reasonable to expect, coming back to discuss the cavitation regime in 
Fig.~\ref{Cd_vs_Cn2}, that Reynolds number effects come into play also where cavitation is occurring so that 
the linear behavior in terms of $C_n^{1/2}$ of 
\eref{eqn:nurick} is not completely fulfilled. On the other hand, in the non-cavitating regime of Fig.~\ref{Cd_vs_Cn2}, 
the reduction of $C_d$ with $C_{n}$ is mainly triggered by the reduction of $Re$, since 
$C_n$ decreases and $C_d$ consistently increases with increasing $Re$.

We have seen that cavitation and Reynolds number  affect the mass flow rate and therefore $C_d$. It could be instructive to consider the above results also looking at the behavior of the velocity and density profiles
and  analyzing how they are affected by cavitation and $Re$.
This will also help to understand
the role of other factors like  compressibility on the behavior of the mass flow rate.

%
%
%
\begin{figure}[t]
\includegraphics[scale=0.4]{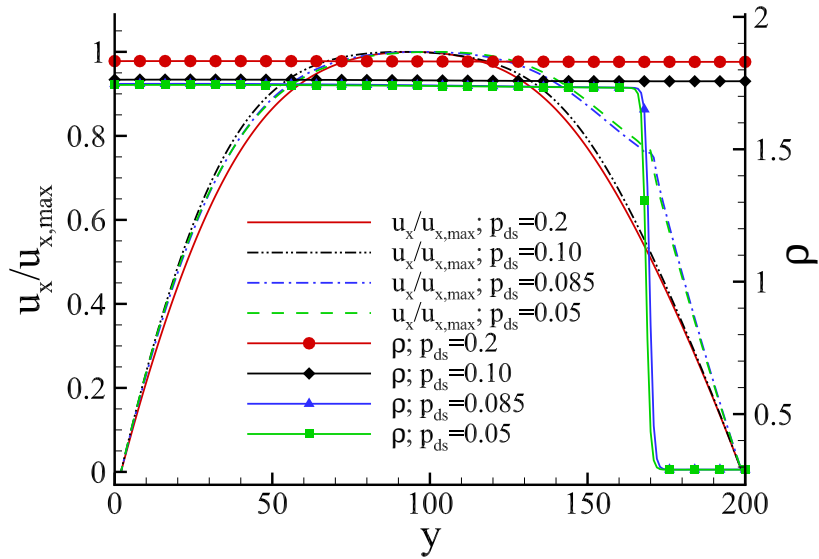}
\caption{Normalized longitudinal velocity profiles along with the density profiles at the position $x=\frac{2}{3}L_x$ across the channel. Simulation parameters: lattice of $601 \times 401$ $\text{lu}$ with fixed-pressure BC at $\theta=0.85$, $\kappa = 0.1$, total inlet pressure $p_{\text{\text{us}}}=0.4$, kinematic viscosity equal to $0.6$ and downstream pressure $p_{\text{ds}} = 0.2$, $0.10$, $0.085$, and $0.05$.}
\label{v_and_rho_profiles}
\end{figure}

Fig.~\ref{v_and_rho_profiles} shows the normalized longitudinal velocity profiles along with the density profiles across the constrained channel at half of its length ($x=\frac{2}{3}L_x$) for the cases with kinematic viscosity $\nu=0.6$ and downstream pressure $p_{\text{ds}}=~0.2, 0.1, 0.085, 0.05$.
The figure shows that density is constant across the channel for the non-cavitating cases with downstream pressure equal to $0.2$ and $0.1$, whereas in the event of cavitation there is a significant reduction of the effective liquid flow cross section and therefore a reduction of $C_d$. The smaller effective liquid flow cross section is caused by a sharp density discontinuity which separates the liquid from the vapor.
It is also visible that, by going from $p_{\text{ds}}=0.2$ to $p_{\text{ds}}=0.1$, compressibility is not negligible since there is an appreciable reduction of the density which obviously affects mass flow rate and therefore $C_d$. 

At the considered longitudinal distance the transversal velocity is negligible and it is at most $2\%$ of the longitudinal one. 
For the non cavitating cases at $p_{\text{ds}}=0.2, 0.10$, see the behaviour of the densities, the shape of the velocity profiles is very close to that of a parabola since the flow is laminar, even if  they are not perfectly symmetric due to the geometry used. A larger $Re$ due to smaller $p_{\text{ds}}$ causes a wider velocity profile which in turn increases the effective flow cross section and consequently $C_d$.
On the other hand, when cavitation occurs, i.e. the cases with $p_{\text{ds}}$ equal to $0.085$ and $0.05$,  the velocity shows a typical linear shear flow profile in the vapor region which persists, with a different slope, moving towards the inner of the system. For both the cavitating cases considered here the ratio between the vapor and the liquid velocity slopes, across the interface, is roughly $4.3$ and it can be compared with the theoretically expected value \cite{papanastasiou-1999} in a system with two phases of different density separated by a horizontal interface under shear flow. It is given by the ratio between liquid and vapor density which here is approximately $5.9$.

A way to evaluate $Re$ effects is also  looking at the boundary layer thickness which is significant to have an idea about the reduction of the effective flow cross section due to viscous forces. Smaller boundary layers means larger effective flow cross section and therefore larger $C_d$. Before the onset of cavitation the presence of the boundary layer in addition to compressibility effects are the main contributions to the discharge coefficient reduction. 
For the non-cavitating cases we can quantify the boundary layer by using the notion of displacement thickness \cite{schlichting-1979}, here defined as
\begin{equation}
\delta^*= h - \int\limits_{0}^{h}\frac{u_x(y)}{u_{x,max}}dy,
\end{equation}
where $u_{x,\text{max}}$ is the maximum value of the longitudinal velocity profile $u_x(y)$.
These are equal to $57.0$~$\text{lu}$ and $53.3$~$\text{lu}$ when the downstream pressure is $0.2$ and $0.1$ respectively, thus showing that by increasing $Re$ the boundary layer decreases and therefore $C_d$ increases.
This analysis indicates that the mass flow rate and therefore the discharge coefficient are affected by different factors. Cavitation causes the most significant reduction of the effective flow cross section. It is further reduced by the boundary layer, which decreases with the Reynolds number. Finally, the compressibility causes a density reduction which in turn reduces the discharge coefficient. This compressibility effect is, however, quite small compared to the previous two.

\section{Conclusions}
We conducted an in-depth numerical investigation of the formation of vapor in a constrained liquid flow using an enhanced forcing-term based lattice Boltzmann model. To model the liquid-vapor transition a free-energy approach based on the van der Waals equation of state was employed. The geometry of the system is a simple two-dimensional channel with a sack-wall obstacle positioned at the top of the channel and well behind the inlet layer inside the system. The channel is thus constrained to half its original width. Two different sets of inflow and outflow boundary conditions were implemented. One where density and velocity at the inlet layer and only the density at the outlet are fixed, and another fixing total pressure at the inflow layer and the static pressure at the outflow layer. The former approach allows for easier control of the velocity profile within the system and was used to investigate the circumstances of local cavitation formation. The latter approach on the other hand better reproduces the situation of experimental work focused on macroscopic hydraulic characterization of cavitation. 

Liquid was forced into the channel from the left to flow past the obstacle. At a sufficient inlet velocity or pressure difference between inlet and outlet boundaries vapor formation underneath the obstacle was observed.

Careful investigation of the local circumstances of the formation of the vapor cavity indicates that it is necessary to reach the liquid branch spinodal pressure. The obtained data imply that this is a necessary but not sufficient condition. The density goes well below the liquid branch spinodal value without inducing the formation of a vapor bubble. 
Our observations then indicate that additional factors, such as viscous stress, interfacial contributions to the local pressure, and the Laplace pressure, are relevant to the opening of a vapor cavity. This is well described by an appropriate generalization of Joseph's \cite{joseph-1998} criterion that includes these contributions. Cavitation occurs when the magnitude of the maximum stress eigenvalue reaches a threshold given by the spinodal pressure corrected by the Laplace contribution. 
The results are in agreement with those of Falcucci~{\it et al.}~\cite{falcucci-2013} but we stress they are obtained from a LBM implementation based on a different thermodynamic model.

This investigation is also the first application of the LBM to the engineering practice of characterizing cavitation through a macroscopic approach by analyzing the behavior of stationary mass flow rate versus pressure drop and discharge coefficient versus both cavitation and Reynolds number. The outcomes found here are qualitatively in good agreement with the experimental ones of Ref.~\onlinecite{payri-2013} and indicate that the mass flow rate and therefore the discharge coefficient are affected by cavitation but also by other factors such as boundary layer and compressibility. 
As theory predicts, a linear increase of the mass flow rate with the square root of the pressure drop was found. However, the occurrence of cavitation causes a reduction in the mass flow growth rate and eventually the system collapses into a choked flow state. Reduction of the mass flow growth rate coincides with a smaller discharge coefficient which shows two different behaviors in the non-cavitating and cavitating regimes.
In the non-cavitating regime, the discharge coefficient grows with the Reynolds number since the reduction of the boundary layer with $Re$ causes larger effective flow cross section. In the case of cavitation, however, vapor causes a significant reduction of the effective liquid flow cross section and therefore a reduction of the discharge coefficient. In this regime, as predicted by Nurick\cite{nurick-1976}, the discharge coefficient grows with the cavitation number $C_{n}$. However, agreement with Nurick law is, at this point, qualitative, since we found the predicted square-root dependence of the discharge coefficient with $C_{n}$ only for a specific value of viscosity. We suspect this is due to Reynolds number effects not taken into account in Nurick's one-dimensional model.

For future research we will investigate the impact of wetting properties of the walls on the cavity formation process and the morphology of cavities.

\acknowledgments
We warmly thank Giacomo Falcucci for detailed discussion of his work.

\appendix

\section{Boundary conditions}
We outline here some details of the implementation of the numerical model with respect to boundary conditions. 

\label{app:bc}
\subsection{Inlet and Outlet Boundary Conditions}
Two different sets of inflow boundary conditions (BC) are employed in this study. The first one is called fixed-density BC and the second one is referred to as fixed-pressure BC. For the fixed-density BC a constant density at the inflow $\rho_{\text{us}}$ and a fixed inflow velocity ${\bf u}_{\text{us}}$ are used. The distribution functions at the inflow layer are then calculated as equilibrium distribution functions according to \eref{eqn:f0} such that $f_{\text{i}} = f_i^0(\rho_{\text{us}}, \vecc{u}_{\text{us}}, \theta)$. Additionally, in order to avoid instabilities at the corners the horizontal inflow velocity is attenuated near the corners according to
\begin{align}
\label{eqn:uxin}
&u_{\text{us},x}(y) = -\frac{u_{\text{us}}}{2} \times \\ \nonumber 
&\lbk \tanh{ \lpar \frac{y - 0.1 L_y}{0.1 L_y}  \rpar } -  \tanh{ \lpar \frac{y-L_y + 0.1 L_y}{0.1 L_y} \rpar } \rbk.
\end{align}
Calculation of the density at the outflow layer follows a similar approach with a zero gradient on the velocity with a fixed outflow density $\rho_{\text{ds}}$. Copying the populations from the layer on the left of the outflow layer and rescaling them by the density, is a simple way of solving this problem:
\beq{eqn:fout}
f_i(x_{\text{ds}}, y) = \frac{\rho_{\text{ds}}}{\rho(x_{\text{ds}}-\Delta x, y)}   f_i(x_{\text{ds}}-\Delta x, y).
\eeq

The fixed-pressure BC are introduced to ensure compatibility with experimental work \cite{payri-2013, nurick-1976, winklhofer-2001} and to obtain the best possible estimate for the cavitation number $C_n$ and its various definitions. For this approach the pressure is fixed at both the inlet and the outlet boundaries.
At the inlet, the total upstream pressure $p_{\text{us}}$ is fixed. The velocity at the inlet in the $x$-direction is calculated from a zero-gradient velocity, i.e. velocities are taken from the layer adjacent to the right of the inlet layer, $u_x (x=0) = u_x(x=\Delta x)$. The $y$-velocity is fixed to vanish $u_y(x=0)=0$. At initialization of the simulation the fluid is at rest and the density is typically chosen to be slightly below the equilibrium liquid density.

At the inlet layer we thus identify the definition of the total pressure with the inlet pressure, i.e. $p_{\text{us}} = p^w + 1/2 \rho \vecc{u}^2$. Replacing $p^w$ with the equation of state of \eref{eqn:vdwEOS} and rearranging the terms we obtain a third order polynomial in $\rho$
\beq{eqn:inflownspol}
0 = \rho^3 - \lpar \frac{4}{9} u^2 + 3 \rpar \rho^2 + \frac{8}{9} \lpar 3 \theta + \frac{3}{2} u^2 - p_{\text{us}} \rpar \rho - \frac{8}{3} p_{\text{us}}.
\eeq
The largest real solution of this expression represents the liquid branch density $\rho_{\text{us}}=\rho_l$ that solves the equation of state for an effective van der Waals pressure of $p_{\text{us}} - 1/2 \rho_l \vecc{u}^2$. This density value is then used to calculate the distribution functions at the boundary layer. This density is streamed to the adjacent layer to the right causing a development of a velocity profile along the x-direction assuming that $p_{\text{us}} > p_{\text{ds}}$.

For the outflow a similar approach is used. The exception is that now only the pressure due to the equation of state $p_{\text{ds}} = p^w(x = L_x)$ enters the boundary condition at the outlet and we have \eref{eqn:inflownspol} but with $u = 0$. Solving again for the largest density fixes $\rho_{\text{ds}} = \rho_l$ at the outflow layer. This density, together with the zero-velocity gradient, is used in \eref{eqn:fout} to determine the distribution functions at the outflow layer.
Including the dynamic pressure contribution in the inlet but not in the outlet pressure may seem counter-intuitive but this is done specifically to reproduce the upstream and downstream pressure definitions of the experimental work in Ref.~\onlinecite{nurick-1976}. Thus the corresponding subsequent definition of the cavitation number $C_n$ in \eref{eqn:CN2} also taken from\cite{nurick-1976} is maintained. 

The effect of the fixed-pressure BC at the outlet is not very significant in the current configuration. However, at the inlet, unlike the fixed profile of \eref{eqn:uxin} in the fixed-density case, the velocity is no longer fixed. Instead the velocity may vary along the inflow layer and even backflow can occur. This has three practical effects for the numerical study: The simulation tends to be less stable during initial formation of the flow profile, to obtain cavitation mean pressure at the inlet layer needs to be chosen higher than in the fixed-density case, and flow profiles and vapor patterns look slightly different.

\subsection{Wall Boundary Conditions}
For the D$2$Q$9$ simulation we give here explicit examples for the wall boundary condition implementation for all cases: flat wall, convex corner, and concave corner. In this implementation the base velocity set is given by $\vecc{v}_{i} = \Delta x / \Delta t \times \{ \{0,0\},$ $\{1,0\},$ $\{0,1\},$ $\{-1,0\},$ $\{0,-1\},$ $\{1,1\},$ $\{-1,1\},$ $\{-1,-1\},$ $\{1,-1\} \}$ for $i=0,..., 8$, respectively.

Let $f_{pi}$ be the outgoing distribution functions 
in a wall lattice site at time $t-\Delta t$ and $f_i$ be those ones
streamed from neighboring sites at time $t$.
For the bottom wall the local fluid density is then given by 
\beq{eqn:bcmas}
\rho = f_{p0}+f_{p7}+f_{p4}+f_{p8}+f_1+f_3+f_8+f_4+f_7.
\eeq
Mass and momentum conservation require
\begin{align}
\label{eqn:bccons1}
\rho & = \sum_{i=0}^8 f_i \\
\label{eqn:bccons2}
\rho u_x & = f_1 + f_5 + f_8 - f_3 - f_6 - f_7 - \frac{\Delta t}{2} F_x \\
\label{eqn:bccons3}
\rho u_y & = f_2 + f_5 + f_6 - f_4 - f_7 - f_8 - \frac{\Delta t}{2} F_y 
\end{align}
which can be solved for the missing populations if one assumes the bounce-back 
prescription $f_2=f_4$ for the population normal to the wall. It then results in
\begin{align}
\label{eqn:bcunknowns1}
f_5 & = f_7 + \frac{1}{2} (f_3 - f_1) - \frac{\Delta t}{4}(F_x+F_y) + \frac{\rho}{2} (u_{w,x} + u_{w,y}), \\
\label{eqn:bcunknowns2}
f_6 & = f_8 + \frac{1}{2} (f_1 - f_3) + \frac{\Delta t}{4}(F_x-F_y) - \frac{\rho}{2} (u_{w,x} - u_{w,y}), \\
\label{eqn:bcunknowns3}
f_0 & = \rho - (f_1 + f_3) - 2 (f_4+f_7+f_8) + \frac{\Delta t}{2} F_y - \rho u_y.
\end{align}
The wall velocity $\vecc{u}_w$ is included here to illustrate how to introduce shear.

In the case of the obstacle-wall concave angle, here on the example of a corner with walls at the top and to the right, the local fluid density is given by 
\beq{eqn:concave}
\rho = f_{p0}+f_{p8}+f_{p1}+f_{p2}+f_{p6}+f_{p5}+f_1+f_2+f_5.
\eeq
Mass and momentum conservation laws and bounce back rules ($f_4=f_2$ and $f_3=f_1$) 
give the unknown distribution functions
\begin{align}
f_7 & = f_5 +\frac{\Delta t}{4}(F_x+F_y), \\
f_6 & = \frac{1}{2} (\rho-f_0) - f_1 - f_2 - f_5 - \frac{\Delta t}{4}F_y, \\
f_8 & = \frac{1}{2} (\rho-f_0) - f_1 - f_2 - f_5 - \frac{\Delta t}{4}F_x.
\end{align}

In the case of the obstacle convex angle the local fluid density 
is given by 
\beq{eqn:convex}
\rho = f_{p0}+f_{p5}+f_1+f_2+f_3+f_4+f_5+f_6+f_8.
\eeq
Mass and momentum conservation gives the unknown distribution functions
\begin{align}
f_7 & = f_5 + \frac{1}{2} (f_1 + f_2) - \frac{1}{2} (f_3 + f_4)
+\frac{\Delta t}{4}(F_x+F_y), \\
f_0 & = \rho-f_1-f_2-f_3-f_4-f_5-f_6-f_7-f_8.
\end{align}

In order to ensure a higher isotropy and reduce spurious velocities at interfaces, which are unavoidable also in our model (see Ref.~\onlinecite{coclite-2014} for details), we calculated the spatial derivatives in (\ref{eqn:force}) by using a general 9-point second-order finite difference scheme (see Refs.~\onlinecite{shan-2006b,sbragaglia-2007} for details).

\section{Spinodal decomposition under shear}
\label{sec:spindecompshear}

\begin{table}
\begin{tabular}{c|c|c|c|c}
$\dot{\gamma}$ & $\kappa$ & velocity field & $\rho_{\text{init}}$ & $t_{\text{sep}}$ \\
\hline
$0.00$ & $0.0$ & free & $1.3875$ & $2 \times 10^4$ \\
 $0.00$ & $0.1$ & free & $1.3875$ & $2 \times 10^4$ \\
$0.02$ & $0.0$ & free & $1.385$  & $1 \times 10^4$ \\
$0.02$ & $0.1$ & free & $1.385$  & $4 \times 10^4$ \\
$0.04$ & $0.0$ & free & $1.385$  & $1 \times 10^4$ \\
$0.04$ & $0.1$ & free & $1.385$  & $5 \times 10^4$ \\
$0.00$ & $0.0$ & fixed & $1.385$  & $3 \times 10^5$ \\
$0.00$ & $0.1$ & fixed & $1.385$  & $>5 \times 10^5$ \\
$0.02$ & $0.0$ & fixed & $1.385$  & $7 \times 10^5$ \\
$0.02$ & $0.1$ & fixed & $1.385$  & $>7 \times 10^5$ \\
$0.04$ & $0.0$ & fixed & $1.385$  & $3 \times 10^5$ \\
$0.04$ & $0.1$ & fixed & $1.385$  & $1 \times 10^6$ \\
\end{tabular}
\caption{Tests of spinodal decomposition under shear in a channel of width $h=20 lu$ at $\theta = 0.9$, $\kappa = 0.0, 0.1$, and shear rates between  $\dot{\gamma} = 0.0$ and $\dot{\gamma} = 0.04$. Indicated are also the type of shear implementation in the velocity field colum, the highest values of the mean initial density $\rho_{init}$ at which phase separation was observed, and the corresponding time $t_{sep}$ after which liquid and vapor were noted.}
\label{tab:shear}
\end{table}

The nucleation process responsible for the formation of vapor shown in this paper is confined to a relatively small region at the inlet of the constrained channel. This region is subjected to shear. It is known that phase transition behavior may change if the material is sheared\cite{wagner-1999, onuki-2002, gonnella-1998, gonnella-1999}. Thus it is prudent to investigate whether the shear is relevant in the ranges of the parameters where cavitation was observed in measurements presented in this paper. We evaluate if and how shear impacts the spinodal density of the fluid. In the simulation discussed in section IV the shear rates underneath the obstacle near the cavity inception point are found to be $0.015 < \dot{\gamma} < 0.020$. To emulate comparable circumstances we set $\dot{\gamma} = 0.02$ and to see the effect of increased shear $\dot{\gamma} = 0.04$. Two simulation scenarios were prepared. The first one, abbreviated as 'free' in table \ref{tab:shear}, is a channel of width $h$ with periodic boundaries in the x-direction and on-grid bounceback on the upper and lower channel walls. These wall boundaries are constructed according to Eqs.~(\ref{eqn:bcunknowns1}) - (\ref{eqn:bcunknowns3}). The lower wall then is assigned a wall velocity $u_{w,x}(y=0) = -\dot{\gamma} h/2$, and the upper wall $u_{w,x}(y=h) = \dot{\gamma} h/2$ thus ensuring in stationary conditions an average shear rate of $\dot{\gamma}$. The second type of preparation, labeled 'fixed', refers to a configuration where in addition to the boundary walls the fluid velocities are constrainted according to the imposed linear shear profile at the time of each collision. These velocities enter the equilibrium distribution functions in the collision term \eref{eqn:LBE1}. Material can still move through advected differences between the local and equilibrium distribution functions but the phase separation dynamics is observed to be slower. The simulation is then initialized with a mean density $\rho_{\text{init}}$ with a very small random offset per lattice node of amplitude of $\Lambda \times 10^{-6}$ where $\Lambda$ is a random number chosen from a flat distribution with zero mean and variance of $0.5$. The initial density is very slightly below the spinodal density measured in simulations of table~\ref{tab:spindata}. Then, if no phase separation is observed after $10^7$ simulation updates, the experiment is repeated at a different initial density $\rho_{\text{init}}$ which is $0.0025$ lower than the previous one.
In order to reproduce the conditions at which cavitation formation is observed underneath the obstacle, we choose a channel width of $h=20lu$. 
This is large enough to avoid interfacial effects (the interface width is $\sim 5 lu$, see, for example, Ref.~\onlinecite{wagner-2006}) and it is comparable to the width of the boundary layer which is $\sim 50lu$ in the right end of the channel (see Figs.~\ref{fig:pmix02500} and \ref{v_and_rho_profiles}) and is less wide near the inlet of the constrained channel. The results of these measurements are given in table \ref{tab:shear}, where the highest initial densities $\rho_{\text{init}}$ for which phase separation is observed, are reported. The observed difference between the measured spinodal density of table~\ref{tab:spindata} and the largest effective density where the system phase separates at $\dot{\gamma} \ne 0$ is $<0.005$. This difference is significantly smaller than the one between analytic spinodal density $1.3916$ and the density $1.315$ at cavitation inception which is $0.08$.


\begin{thebibliography}{74}%
\makeatletter
\providecommand \@ifxundefined [1]{%
 \@ifx{#1\undefined}
}%
\providecommand \@ifnum [1]{%
 \ifnum #1\expandafter \@firstoftwo
 \else \expandafter \@secondoftwo
 \fi
}%
\providecommand \@ifx [1]{%
 \ifx #1\expandafter \@firstoftwo
 \else \expandafter \@secondoftwo
 \fi
}%
\providecommand \natexlab [1]{#1}%
\providecommand \enquote  [1]{``#1''}%
\providecommand \bibnamefont  [1]{#1}%
\providecommand \bibfnamefont [1]{#1}%
\providecommand \citenamefont [1]{#1}%
\providecommand \href@noop [0]{\@secondoftwo}%
\providecommand \href [0]{\begingroup \@sanitize@url \@href}%
\providecommand \@href[1]{\@@startlink{#1}\@@href}%
\providecommand \@@href[1]{\endgroup#1\@@endlink}%
\providecommand \@sanitize@url [0]{\catcode `\\12\catcode `\$12\catcode
  `\&12\catcode `\#12\catcode `\^12\catcode `\_12\catcode `\%12\relax}%
\providecommand \@@startlink[1]{}%
\providecommand \@@endlink[0]{}%
\providecommand \url  [0]{\begingroup\@sanitize@url \@url }%
\providecommand \@url [1]{\endgroup\@href {#1}{\urlprefix }}%
\providecommand \urlprefix  [0]{URL }%
\providecommand \Eprint [0]{\href }%
\providecommand \doibase [0]{http://dx.doi.org/}%
\providecommand \selectlanguage [0]{\@gobble}%
\providecommand \bibinfo  [0]{\@secondoftwo}%
\providecommand \bibfield  [0]{\@secondoftwo}%
\providecommand \translation [1]{[#1]}%
\providecommand \BibitemOpen [0]{}%
\providecommand \bibitemStop [0]{}%
\providecommand \bibitemNoStop [0]{.\EOS\space}%
\providecommand \EOS [0]{\spacefactor3000\relax}%
\providecommand \BibitemShut  [1]{\csname bibitem#1\endcsname}%
\let\auto@bib@innerbib\@empty
\bibitem [{\citenamefont {Brennen}(1995)}]{brennen-1995}%
  \BibitemOpen
  \bibfield  {author} {\bibinfo {author} {\bibfnamefont {C.~E.}\ \bibnamefont
  {Brennen}},\ }\href@noop {} {\emph {\bibinfo {title} {Cavitation and Bubble
  Dynamics}}}\ (\bibinfo  {publisher} {Oxford University Press},\ \bibinfo
  {year} {1995})\BibitemShut {NoStop}%
\bibitem [{\citenamefont {Dabiri}, \citenamefont {Sirignano},\ and\
  \citenamefont {Joseph}(2007)}]{dabiri-2007}%
  \BibitemOpen
  \bibfield  {author} {\bibinfo {author} {\bibfnamefont {S.}~\bibnamefont
  {Dabiri}}, \bibinfo {author} {\bibfnamefont {W.~A.}\ \bibnamefont
  {Sirignano}}, \ and\ \bibinfo {author} {\bibfnamefont {D.~D.}\ \bibnamefont
  {Joseph}},\ }\bibfield  {title} {\enquote {\bibinfo {title} {Cavitation in an
  orifice flow},}\ }\href@noop {} {\bibfield  {journal} {\bibinfo  {journal}
  {Physics of Fluids}\ }\textbf {\bibinfo {volume} {19}},\ \bibinfo {pages}
  {072112} (\bibinfo {year} {2007})}\BibitemShut {NoStop}%
\bibitem [{\citenamefont {Koivula}()}]{koivula-2000}%
  \BibitemOpen
  \bibfield  {author} {\bibinfo {author} {\bibfnamefont {T.}~\bibnamefont
  {Koivula}},\ }\bibfield  {title} {\enquote {\bibinfo {title} {On cavitation
  in fluid power},}\ }in\ \href@noop {} {\emph {\bibinfo {booktitle}
  {Proceedings of the 1st FPNI PhD Symposium, Hamburg, Germany, September
  20{\textendash}22, 2000}}},\ \bibinfo {editor} {edited by\ \bibinfo {editor}
  {\bibfnamefont {M.}~\bibnamefont {Ivantysynova}}}\ (\bibinfo  {publisher}
  {Technical University of Hamburg-Harburg})\ pp.\ \bibinfo {pages}
  {371--382}\BibitemShut {NoStop}%
\bibitem [{\citenamefont {Lohse}(2005)}]{lohse-2005}%
  \BibitemOpen
  \bibfield  {author} {\bibinfo {author} {\bibfnamefont {D.}~\bibnamefont
  {Lohse}},\ }\bibfield  {title} {\enquote {\bibinfo {title} {{Cavitation hots
  up}},}\ }\href@noop {} {\bibfield  {journal} {\bibinfo  {journal} {Nature}\
  }\textbf {\bibinfo {volume} {434}},\ \bibinfo {pages} {33} (\bibinfo {year}
  {2005})}\BibitemShut {NoStop}%
\bibitem [{\citenamefont {Reynolds}(1873)}]{reynolds-1873}%
  \BibitemOpen
  \bibfield  {author} {\bibinfo {author} {\bibfnamefont {O.}~\bibnamefont
  {Reynolds}},\ }\bibfield  {title} {\enquote {\bibinfo {title} {The causes of
  the racing of the engines of screw steamers investigated theoretically and by
  experiment},}\ }\href@noop {} {\bibfield  {journal} {\bibinfo  {journal}
  {Transactions of the Royal Institution of Naval Architects}\ }\textbf
  {\bibinfo {volume} {14}},\ \bibinfo {pages} {56} (\bibinfo {year}
  {1873})}\BibitemShut {NoStop}%
\bibitem [{\citenamefont {Parsons}(1906)}]{parsons-1906}%
  \BibitemOpen
  \bibfield  {author} {\bibinfo {author} {\bibfnamefont {C.~A.}\ \bibnamefont
  {Parsons}},\ }\bibfield  {title} {\enquote {\bibinfo {title} {{The steam
  turbine on land and at sea}},}\ }\href@noop {} {\bibfield  {journal}
  {\bibinfo  {journal} {Lecture to the Royal Institution, London}\ } (\bibinfo
  {year} {1906})}\BibitemShut {NoStop}%
\bibitem [{\citenamefont {Kawanami}\ \emph {et~al.}(1997)\citenamefont
  {Kawanami}, \citenamefont {Kato}, \citenamefont {Yamaguchi}, \citenamefont
  {Tanimura},\ and\ \citenamefont {Tagaya}}]{kawanami-1997}%
  \BibitemOpen
  \bibfield  {author} {\bibinfo {author} {\bibfnamefont {Y.}~\bibnamefont
  {Kawanami}}, \bibinfo {author} {\bibfnamefont {H.}~\bibnamefont {Kato}},
  \bibinfo {author} {\bibfnamefont {H.}~\bibnamefont {Yamaguchi}}, \bibinfo
  {author} {\bibfnamefont {M.}~\bibnamefont {Tanimura}}, \ and\ \bibinfo
  {author} {\bibfnamefont {Y.}~\bibnamefont {Tagaya}},\ }\bibfield  {title}
  {\enquote {\bibinfo {title} {{Mechanism and Control of Cloud Cavitation}},}\
  }\href@noop {} {\bibfield  {journal} {\bibinfo  {journal} {Journal of Fluids
  Engineering}\ }\textbf {\bibinfo {volume} {119}},\ \bibinfo {pages} {788}
  (\bibinfo {year} {1997})}\BibitemShut {NoStop}%
\bibitem [{\citenamefont {Weiland}\ and\ \citenamefont
  {Vlachos}(2012)}]{weiland-2012}%
  \BibitemOpen
  \bibfield  {author} {\bibinfo {author} {\bibfnamefont {C.}~\bibnamefont
  {Weiland}}\ and\ \bibinfo {author} {\bibfnamefont {P.~P.}\ \bibnamefont
  {Vlachos}},\ }\bibfield  {title} {\enquote {\bibinfo {title} {{Time-scale for
  critical growth of partial and supercavitation development over impulsively
  translating projectiles}},}\ }\href@noop {} {\bibfield  {journal} {\bibinfo
  {journal} {International Journal of Multiphase Flow}\ }\textbf {\bibinfo
  {volume} {38}},\ \bibinfo {pages} {73} (\bibinfo {year} {2012})}\BibitemShut
  {NoStop}%
\bibitem [{\citenamefont {Huang}, \citenamefont {Ducoin},\ and\ \citenamefont
  {Young}(2013)}]{huang-2013}%
  \BibitemOpen
  \bibfield  {author} {\bibinfo {author} {\bibfnamefont {B.}~\bibnamefont
  {Huang}}, \bibinfo {author} {\bibfnamefont {A.}~\bibnamefont {Ducoin}}, \
  and\ \bibinfo {author} {\bibfnamefont {Y.~L.}\ \bibnamefont {Young}},\
  }\bibfield  {title} {\enquote {\bibinfo {title} {{Physical and numerical
  investigation of cavitating flows around a pitching hydrofoil}},}\
  }\href@noop {} {\bibfield  {journal} {\bibinfo  {journal} {Physics of
  Fluids}\ }\textbf {\bibinfo {volume} {25}},\ \bibinfo {pages} {102109}
  (\bibinfo {year} {2013})}\BibitemShut {NoStop}%
\bibitem [{\citenamefont {Winklhofer}\ \emph {et~al.}()\citenamefont
  {Winklhofer}, \citenamefont {Kull}, \citenamefont {Kelz},\ and\ \citenamefont
  {Morozov}}]{winklhofer-2001}%
  \BibitemOpen
  \bibfield  {author} {\bibinfo {author} {\bibfnamefont {E.}~\bibnamefont
  {Winklhofer}}, \bibinfo {author} {\bibfnamefont {E.}~\bibnamefont {Kull}},
  \bibinfo {author} {\bibfnamefont {E.}~\bibnamefont {Kelz}}, \ and\ \bibinfo
  {author} {\bibfnamefont {A.}~\bibnamefont {Morozov}},\ }\bibfield  {title}
  {\enquote {\bibinfo {title} {Comprehensive hydraulic and flow field
  documentation in model throttle experiments under cavitation conditions},}\
  }in\ \href@noop {} {\emph {\bibinfo {booktitle} {Proceedings 17th ANNUAL
  CONFERENCE ON LIQUID ATOMIZATION \& SPRAY SYSTEMS, Zurich, Switzerland,
  September 2{\textendash}6, 2001}}},\ \bibinfo {editor} {edited by\ \bibinfo
  {editor} {\bibfnamefont {B.}~\bibnamefont {Ineichen}}}\ (\bibinfo
  {publisher} {I.C. Engines and Combustion Laboratory at ETH,
  Zurich})\BibitemShut {NoStop}%
\bibitem [{\citenamefont {Payri}\ \emph {et~al.}(2013)\citenamefont {Payri},
  \citenamefont {Salvador}, \citenamefont {Gimeno},\ and\ \citenamefont
  {Venegas}}]{payri-2013}%
  \BibitemOpen
  \bibfield  {author} {\bibinfo {author} {\bibfnamefont {R.}~\bibnamefont
  {Payri}}, \bibinfo {author} {\bibfnamefont {F.~J.}\ \bibnamefont {Salvador}},
  \bibinfo {author} {\bibfnamefont {J.}~\bibnamefont {Gimeno}}, \ and\ \bibinfo
  {author} {\bibfnamefont {O.}~\bibnamefont {Venegas}},\ }\bibfield  {title}
  {\enquote {\bibinfo {title} {{Study of cavitation phenomenon using different
  fuels in a transparent nozzle by hydraulic characterization and
  visualization}},}\ }\href@noop {} {\bibfield  {journal} {\bibinfo  {journal}
  {Experimental Thermal and Fluid Science}\ }\textbf {\bibinfo {volume} {44}},\
  \bibinfo {pages} {235} (\bibinfo {year} {2013})}\BibitemShut {NoStop}%
\bibitem [{\citenamefont {Sou}, \citenamefont {Hosokawa},\ and\ \citenamefont
  {Tomiyama}(2007)}]{suo-2007}%
  \BibitemOpen
  \bibfield  {author} {\bibinfo {author} {\bibfnamefont {A.}~\bibnamefont
  {Sou}}, \bibinfo {author} {\bibfnamefont {S.}~\bibnamefont {Hosokawa}}, \
  and\ \bibinfo {author} {\bibfnamefont {A.}~\bibnamefont {Tomiyama}},\
  }\bibfield  {title} {\enquote {\bibinfo {title} {{Effects of cavitation in a
  nozzle on liquid jet atomization}},}\ }\href@noop {} {\bibfield  {journal}
  {\bibinfo  {journal} {International Journal of Heat and Mass Transfer}\
  }\textbf {\bibinfo {volume} {50}},\ \bibinfo {pages} {3575} (\bibinfo {year}
  {2007})}\BibitemShut {NoStop}%
\bibitem [{\citenamefont {Sou}\ \emph {et~al.}(2006)\citenamefont {Sou},
  \citenamefont {Tomiyama}, \citenamefont {Hosokawa}, \citenamefont
  {Nigorikawa},\ and\ \citenamefont {Maeda}}]{suo-2006}%
  \BibitemOpen
  \bibfield  {author} {\bibinfo {author} {\bibfnamefont {A.}~\bibnamefont
  {Sou}}, \bibinfo {author} {\bibfnamefont {A.}~\bibnamefont {Tomiyama}},
  \bibinfo {author} {\bibfnamefont {S.}~\bibnamefont {Hosokawa}}, \bibinfo
  {author} {\bibfnamefont {S.}~\bibnamefont {Nigorikawa}}, \ and\ \bibinfo
  {author} {\bibfnamefont {T.}~\bibnamefont {Maeda}},\ }\bibfield  {title}
  {\enquote {\bibinfo {title} {{Cavitation in a two-dimensional nozzle and
  liquid jet atomization (LDV measurement of liquid velocity in a nozzle)}},}\
  }\href@noop {} {\bibfield  {journal} {\bibinfo  {journal} {JSME International
  Journal Series B}\ }\textbf {\bibinfo {volume} {49}},\ \bibinfo {pages}
  {1253} (\bibinfo {year} {2006})}\BibitemShut {NoStop}%
\bibitem [{\citenamefont {Mishra}\ and\ \citenamefont
  {Peles}(2005)}]{mishra-2005}%
  \BibitemOpen
  \bibfield  {author} {\bibinfo {author} {\bibfnamefont {C.}~\bibnamefont
  {Mishra}}\ and\ \bibinfo {author} {\bibfnamefont {Y.}~\bibnamefont {Peles}},\
  }\bibfield  {title} {\enquote {\bibinfo {title} {{Cavitation in flow through
  a micro-orifice inside a silicon microchannel}},}\ }\href@noop {} {\bibfield
  {journal} {\bibinfo  {journal} {Physics of Fluids}\ }\textbf {\bibinfo
  {volume} {17}},\ \bibinfo {pages} {013601} (\bibinfo {year}
  {2005})}\BibitemShut {NoStop}%
\bibitem [{\citenamefont {Tamaki}\ \emph {et~al.}(1998)\citenamefont {Tamaki},
  \citenamefont {Shimizu}, \citenamefont {Nishida},\ and\ \citenamefont
  {Hiroyasu}}]{tamaki-1998}%
  \BibitemOpen
  \bibfield  {author} {\bibinfo {author} {\bibfnamefont {N.}~\bibnamefont
  {Tamaki}}, \bibinfo {author} {\bibfnamefont {M.}~\bibnamefont {Shimizu}},
  \bibinfo {author} {\bibfnamefont {K.}~\bibnamefont {Nishida}}, \ and\
  \bibinfo {author} {\bibfnamefont {H.}~\bibnamefont {Hiroyasu}},\ }\bibfield
  {title} {\enquote {\bibinfo {title} {{Effects of cavitation and internal flow
  on atomization of a liquid jet}},}\ }\href@noop {} {\bibfield  {journal}
  {\bibinfo  {journal} {Atomization Sprays}\ }\textbf {\bibinfo {volume} {8}},\
  \bibinfo {pages} {179} (\bibinfo {year} {1998})}\BibitemShut {NoStop}%
\bibitem [{\citenamefont {Hiroyasu}(2000)}]{hiroyasu-2000}%
  \BibitemOpen
  \bibfield  {author} {\bibinfo {author} {\bibfnamefont {H.}~\bibnamefont
  {Hiroyasu}},\ }\bibfield  {title} {\enquote {\bibinfo {title} {{Spray breakup
  mechanism from the hole-type nozzle and its applications}},}\ }\href@noop {}
  {\bibfield  {journal} {\bibinfo  {journal} {Atomization Sprays}\ }\textbf
  {\bibinfo {volume} {10}},\ \bibinfo {pages} {511} (\bibinfo {year}
  {2000})}\BibitemShut {NoStop}%
\bibitem [{\citenamefont {Payri}\ \emph {et~al.}(2004)\citenamefont {Payri},
  \citenamefont {Berm\'udez}, \citenamefont {Payri},\ and\ \citenamefont
  {Salvador}}]{payri-2004}%
  \BibitemOpen
  \bibfield  {author} {\bibinfo {author} {\bibfnamefont {F.}~\bibnamefont
  {Payri}}, \bibinfo {author} {\bibfnamefont {V.}~\bibnamefont {Berm\'udez}},
  \bibinfo {author} {\bibfnamefont {R.}~\bibnamefont {Payri}}, \ and\ \bibinfo
  {author} {\bibfnamefont {F.~J.}\ \bibnamefont {Salvador}},\ }\bibfield
  {title} {\enquote {\bibinfo {title} {{The influence of cavitation on the
  internal flow and the spray characteristics in diesel injection nozzles}},}\
  }\href@noop {} {\bibfield  {journal} {\bibinfo  {journal} {Fuel}\ }\textbf
  {\bibinfo {volume} {83}},\ \bibinfo {pages} {419} (\bibinfo {year}
  {2004})}\BibitemShut {NoStop}%
\bibitem [{\citenamefont {{De Giorgi}}, \citenamefont {Ficarella},\ and\
  \citenamefont {Tarantino}(2013)}]{de-giorgi-2013}%
  \BibitemOpen
  \bibfield  {author} {\bibinfo {author} {\bibfnamefont {M.~G.}\ \bibnamefont
  {{De Giorgi}}}, \bibinfo {author} {\bibfnamefont {A.}~\bibnamefont
  {Ficarella}}, \ and\ \bibinfo {author} {\bibfnamefont {M.}~\bibnamefont
  {Tarantino}},\ }\bibfield  {title} {\enquote {\bibinfo {title} {{Evaluating
  cavitation regimes in an internal orifice at different temperatures using
  frequency analysis and visualization}},}\ }\href@noop {} {\bibfield
  {journal} {\bibinfo  {journal} {International Journal of Heat and Fluid
  Flow}\ }\textbf {\bibinfo {volume} {39}},\ \bibinfo {pages} {160} (\bibinfo
  {year} {2013})}\BibitemShut {NoStop}%
\bibitem [{\citenamefont {Gavaises}\ \emph {et~al.}(2009)\citenamefont
  {Gavaises}, \citenamefont {Andriotis}, \citenamefont {Papoulias},
  \citenamefont {Mitroglou},\ and\ \citenamefont
  {Theodorakakos}}]{gavaises-2009}%
  \BibitemOpen
  \bibfield  {author} {\bibinfo {author} {\bibfnamefont {M.}~\bibnamefont
  {Gavaises}}, \bibinfo {author} {\bibfnamefont {A.}~\bibnamefont {Andriotis}},
  \bibinfo {author} {\bibfnamefont {D.}~\bibnamefont {Papoulias}}, \bibinfo
  {author} {\bibfnamefont {N.}~\bibnamefont {Mitroglou}}, \ and\ \bibinfo
  {author} {\bibfnamefont {A.}~\bibnamefont {Theodorakakos}},\ }\bibfield
  {title} {\enquote {\bibinfo {title} {{Characterization of string cavitation
  in large-scale Diesel nozzles with tapered holes}},}\ }\href@noop {}
  {\bibfield  {journal} {\bibinfo  {journal} {Physics of Fluids}\ }\textbf
  {\bibinfo {volume} {21}},\ \bibinfo {pages} {052107} (\bibinfo {year}
  {2009})}\BibitemShut {NoStop}%
\bibitem [{\citenamefont {Giannadakis}, \citenamefont {Gavaises},\ and\
  \citenamefont {Arcoumanis}(2008)}]{giannadakis-2008}%
  \BibitemOpen
  \bibfield  {author} {\bibinfo {author} {\bibfnamefont {E.}~\bibnamefont
  {Giannadakis}}, \bibinfo {author} {\bibfnamefont {M.}~\bibnamefont
  {Gavaises}}, \ and\ \bibinfo {author} {\bibfnamefont {C.}~\bibnamefont
  {Arcoumanis}},\ }\bibfield  {title} {\enquote {\bibinfo {title} {{Modelling
  of cavitation in diesel injector nozzles}},}\ }\href@noop {} {\bibfield
  {journal} {\bibinfo  {journal} {Journal of Fluid Mechanics}\ }\textbf
  {\bibinfo {volume} {616}},\ \bibinfo {pages} {153} (\bibinfo {year}
  {2008})}\BibitemShut {NoStop}%
\bibitem [{\citenamefont {Kuo}\ and\ \citenamefont
  {Acharya}(2012)}]{kuo-acharya-2012}%
  \BibitemOpen
  \bibfield  {author} {\bibinfo {author} {\bibfnamefont {K.~K.}\ \bibnamefont
  {Kuo}}\ and\ \bibinfo {author} {\bibfnamefont {R.}~\bibnamefont {Acharya}},\
  }\href@noop {} {\emph {\bibinfo {title} {Fundamentals of turbulent and
  multiphase combustion}}}\ (\bibinfo  {publisher} {John Wiley \& Sons, Inc.},\
  \bibinfo {year} {2012})\BibitemShut {NoStop}%
\bibitem [{\citenamefont {Chung}, \citenamefont {Pak},\ and\ \citenamefont
  {Chang}(2004)}]{chung-2004}%
  \BibitemOpen
  \bibfield  {author} {\bibinfo {author} {\bibfnamefont {M.~S.}\ \bibnamefont
  {Chung}}, \bibinfo {author} {\bibfnamefont {S.~K.}\ \bibnamefont {Pak}}, \
  and\ \bibinfo {author} {\bibfnamefont {K.~S.}\ \bibnamefont {Chang}},\
  }\bibfield  {title} {\enquote {\bibinfo {title} {A numerical study of
  two-phase flow using a two-dimensional two-fluid model},}\ }\href@noop {}
  {\bibfield  {journal} {\bibinfo  {journal} {Numerical Heat Transfer, Part A}\
  }\textbf {\bibinfo {volume} {45}},\ \bibinfo {pages} {1049} (\bibinfo {year}
  {2004})}\BibitemShut {NoStop}%
\bibitem [{\citenamefont {Niu}\ and\ \citenamefont {Lin}(2006)}]{niu-2006}%
  \BibitemOpen
  \bibfield  {author} {\bibinfo {author} {\bibfnamefont {Y.~Y.}\ \bibnamefont
  {Niu}}\ and\ \bibinfo {author} {\bibfnamefont {Y.~M.}\ \bibnamefont {Lin}},\
  }\bibfield  {title} {\enquote {\bibinfo {title} {A multiscale model to
  predict slurry erosion in cavitated duct flows},}\ }\href@noop {} {\bibfield
  {journal} {\bibinfo  {journal} {Numerical Heat Transfer, Part A}\ }\textbf
  {\bibinfo {volume} {50}},\ \bibinfo {pages} {545} (\bibinfo {year}
  {2006})}\BibitemShut {NoStop}%
\bibitem [{\citenamefont {Darbandi}\ and\ \citenamefont
  {Sadeghi}(2010)}]{darbandi-2010}%
  \BibitemOpen
  \bibfield  {author} {\bibinfo {author} {\bibfnamefont {M.}~\bibnamefont
  {Darbandi}}\ and\ \bibinfo {author} {\bibfnamefont {H.}~\bibnamefont
  {Sadeghi}},\ }\bibfield  {title} {\enquote {\bibinfo {title} {Numerical
  simulation of orifice cavitating flows using two-fluid and three-fluid
  cavitation models},}\ }\href@noop {} {\bibfield  {journal} {\bibinfo
  {journal} {Numerical Heat Transfer, Part A}\ }\textbf {\bibinfo {volume}
  {58}},\ \bibinfo {pages} {505} (\bibinfo {year} {2010})}\BibitemShut
  {NoStop}%
\bibitem [{\citenamefont {Yeom}\ and\ \citenamefont {Chang}(2006)}]{yeom-2006}%
  \BibitemOpen
  \bibfield  {author} {\bibinfo {author} {\bibfnamefont {G.~S.}\ \bibnamefont
  {Yeom}}\ and\ \bibinfo {author} {\bibfnamefont {K.~S.}\ \bibnamefont
  {Chang}},\ }\bibfield  {title} {\enquote {\bibinfo {title} {Numerical
  simulation of two-fluid two-phase flows by hll scheme using an approximate
  jacobian matrix},}\ }\href@noop {} {\bibfield  {journal} {\bibinfo  {journal}
  {Numerical Heat Transfer, Part B}\ }\textbf {\bibinfo {volume} {49}},\
  \bibinfo {pages} {155--177} (\bibinfo {year} {2006})}\BibitemShut {NoStop}%
\bibitem [{\citenamefont {Hirt}\ and\ \citenamefont
  {Nichols}(1981)}]{hirt-1981}%
  \BibitemOpen
  \bibfield  {author} {\bibinfo {author} {\bibfnamefont {C.~W.}\ \bibnamefont
  {Hirt}}\ and\ \bibinfo {author} {\bibfnamefont {B.~D.}\ \bibnamefont
  {Nichols}},\ }\bibfield  {title} {\enquote {\bibinfo {title} {{Volume of
  fluid (VOF) method for the dynamics of free boundaries}},}\ }\href@noop {}
  {\bibfield  {journal} {\bibinfo  {journal} {Journal of Computational
  Physics}\ }\textbf {\bibinfo {volume} {39}},\ \bibinfo {pages} {201}
  (\bibinfo {year} {1981})}\BibitemShut {NoStop}%
\bibitem [{\citenamefont {Wang}\ \emph {et~al.}(2001)\citenamefont {Wang},
  \citenamefont {Senocak}, \citenamefont {Shyy}, \citenamefont {Ikohagi},\ and\
  \citenamefont {Cao}}]{wang-2001}%
  \BibitemOpen
  \bibfield  {author} {\bibinfo {author} {\bibfnamefont {G.}~\bibnamefont
  {Wang}}, \bibinfo {author} {\bibfnamefont {I.}~\bibnamefont {Senocak}},
  \bibinfo {author} {\bibfnamefont {W.}~\bibnamefont {Shyy}}, \bibinfo {author}
  {\bibfnamefont {T.}~\bibnamefont {Ikohagi}}, \ and\ \bibinfo {author}
  {\bibfnamefont {S.}~\bibnamefont {Cao}},\ }\bibfield  {title} {\enquote
  {\bibinfo {title} {{Dynamics of attached turbulent cavitating flows}},}\
  }\href@noop {} {\bibfield  {journal} {\bibinfo  {journal} {Progress in
  Aerospace Sciences}\ }\textbf {\bibinfo {volume} {37}},\ \bibinfo {pages}
  {551} (\bibinfo {year} {2001})}\BibitemShut {NoStop}%
\bibitem [{\citenamefont {Utturkar}\ \emph {et~al.}(2005)\citenamefont
  {Utturkar}, \citenamefont {Wu}, \citenamefont {Wang},\ and\ \citenamefont
  {Shyy}}]{utturkar-2005}%
  \BibitemOpen
  \bibfield  {author} {\bibinfo {author} {\bibfnamefont {Y.}~\bibnamefont
  {Utturkar}}, \bibinfo {author} {\bibfnamefont {J.}~\bibnamefont {Wu}},
  \bibinfo {author} {\bibfnamefont {G.}~\bibnamefont {Wang}}, \ and\ \bibinfo
  {author} {\bibfnamefont {W.}~\bibnamefont {Shyy}},\ }\bibfield  {title}
  {\enquote {\bibinfo {title} {{Recent progress in modeling of cryogenic
  cavitation for liquid rocket propulsion}},}\ }\href@noop {} {\bibfield
  {journal} {\bibinfo  {journal} {Progress in Aerospace Sciences}\ }\textbf
  {\bibinfo {volume} {41}},\ \bibinfo {pages} {558} (\bibinfo {year}
  {2005})}\BibitemShut {NoStop}%
\bibitem [{\citenamefont {Tseng}\ \emph {et~al.}(2010)\citenamefont {Tseng},
  \citenamefont {Wei}, \citenamefont {Wang},\ and\ \citenamefont
  {Shyy}}]{tseng-2010}%
  \BibitemOpen
  \bibfield  {author} {\bibinfo {author} {\bibfnamefont {C.~C.}\ \bibnamefont
  {Tseng}}, \bibinfo {author} {\bibfnamefont {Y.}~\bibnamefont {Wei}}, \bibinfo
  {author} {\bibfnamefont {G.}~\bibnamefont {Wang}}, \ and\ \bibinfo {author}
  {\bibfnamefont {W.}~\bibnamefont {Shyy}},\ }\bibfield  {title} {\enquote
  {\bibinfo {title} {{Modeling of turbulent, isothermal and cryogenic
  cavitation under attached conditions}},}\ }\href@noop {} {\bibfield
  {journal} {\bibinfo  {journal} {Acta Mechanica Sinica}\ }\textbf {\bibinfo
  {volume} {26}},\ \bibinfo {pages} {325} (\bibinfo {year} {2010})}\BibitemShut
  {NoStop}%
\bibitem [{\citenamefont {Salvador}\ \emph {et~al.}(2013)\citenamefont
  {Salvador}, \citenamefont {Mart\`inez-L\'opez}, \citenamefont {Romero},\ and\
  \citenamefont {Rosell\'o}}]{salvador-2013}%
  \BibitemOpen
  \bibfield  {author} {\bibinfo {author} {\bibfnamefont {F.~J.}\ \bibnamefont
  {Salvador}}, \bibinfo {author} {\bibfnamefont {J.}~\bibnamefont
  {Mart\`inez-L\'opez}}, \bibinfo {author} {\bibfnamefont {J.~V.}\ \bibnamefont
  {Romero}}, \ and\ \bibinfo {author} {\bibfnamefont {M.~D.}\ \bibnamefont
  {Rosell\'o}},\ }\bibfield  {title} {\enquote {\bibinfo {title}
  {{Computational study of the cavitation phenomenon and its interaction with
  the turbulence developed in diesel injector nozzles by Large Eddy Simulation
  (LES)}},}\ }\href@noop {} {\bibfield  {journal} {\bibinfo  {journal}
  {Mathematical and Computer Modelling}\ }\textbf {\bibinfo {volume} {57}},\
  \bibinfo {pages} {1656} (\bibinfo {year} {2013})}\BibitemShut {NoStop}%
\bibitem [{\citenamefont {Goncalv\`es}\ and\ \citenamefont
  {Charri\`ere}(2014)}]{goncalves-2014}%
  \BibitemOpen
  \bibfield  {author} {\bibinfo {author} {\bibfnamefont {E.}~\bibnamefont
  {Goncalv\`es}}\ and\ \bibinfo {author} {\bibfnamefont {B.}~\bibnamefont
  {Charri\`ere}},\ }\bibfield  {title} {\enquote {\bibinfo {title} {{Modelling
  for isothermal cavitation with a four-equation model}},}\ }\href@noop {}
  {\bibfield  {journal} {\bibinfo  {journal} {International Journal of
  Multiphase Flow}\ }\textbf {\bibinfo {volume} {59}},\ \bibinfo {pages} {54}
  (\bibinfo {year} {2014})}\BibitemShut {NoStop}%
\bibitem [{\citenamefont {Knapp}, \citenamefont {Daily},\ and\ \citenamefont
  {Hammitt}(1970)}]{knapp-1970}%
  \BibitemOpen
  \bibfield  {author} {\bibinfo {author} {\bibfnamefont {R.~T.}\ \bibnamefont
  {Knapp}}, \bibinfo {author} {\bibfnamefont {J.~W.}\ \bibnamefont {Daily}}, \
  and\ \bibinfo {author} {\bibfnamefont {F.~G.}\ \bibnamefont {Hammitt}},\
  }\href@noop {} {\emph {\bibinfo {title} {Cavitation}}}\ (\bibinfo
  {publisher} {McGraw-Hill},\ \bibinfo {year} {1970})\BibitemShut {NoStop}%
\bibitem [{\citenamefont {Lecoffre}(1999)}]{lecoffre-1999}%
  \BibitemOpen
  \bibfield  {author} {\bibinfo {author} {\bibfnamefont {Y.}~\bibnamefont
  {Lecoffre}},\ }\href@noop {} {\emph {\bibinfo {title} {Cavitation bubble
  trackers}}}\ (\bibinfo  {publisher} {CRC Press},\ \bibinfo {year}
  {1999})\BibitemShut {NoStop}%
\bibitem [{\citenamefont {Joseph.}(1998)}]{joseph-1998}%
  \BibitemOpen
  \bibfield  {author} {\bibinfo {author} {\bibfnamefont {D.~D.}\ \bibnamefont
  {Joseph.}},\ }\bibfield  {title} {\enquote {\bibinfo {title} {{Cavitation and
  the state of stress in a flowing liquid}},}\ }\href@noop {} {\bibfield
  {journal} {\bibinfo  {journal} {Journal of Fluid Mechanics}\ }\textbf
  {\bibinfo {volume} {366}},\ \bibinfo {pages} {367} (\bibinfo {year}
  {1998})}\BibitemShut {NoStop}%
\bibitem [{\citenamefont {Bray}(1994)}]{bray-1994}%
  \BibitemOpen
  \bibfield  {author} {\bibinfo {author} {\bibfnamefont {A.}~\bibnamefont
  {Bray}},\ }\bibfield  {title} {\enquote {\bibinfo {title} {{Theory of
  Phase-Ordering Kinetics}},}\ }\href@noop {} {\bibfield  {journal} {\bibinfo
  {journal} {Adv.~Phys.}\ }\textbf {\bibinfo {volume} {43}},\ \bibinfo {pages}
  {357--459} (\bibinfo {year} {1994})}\BibitemShut {NoStop}%
\bibitem [{\citenamefont {Onuki}(2002)}]{onuki-2002}%
  \BibitemOpen
  \bibfield  {author} {\bibinfo {author} {\bibfnamefont {A.}~\bibnamefont
  {Onuki}},\ }\href@noop {} {\emph {\bibinfo {title} {Phase Transition
  Dynamics}}},\ \bibinfo {edition} {1st}\ ed.\ (\bibinfo  {publisher}
  {Cambridge University Press},\ \bibinfo {year} {2002})\BibitemShut {NoStop}%
\bibitem [{\citenamefont {Succi}(2001)}]{succi-2001}%
  \BibitemOpen
  \bibfield  {author} {\bibinfo {author} {\bibfnamefont {S.}~\bibnamefont
  {Succi}},\ }\href@noop {} {\emph {\bibinfo {title} {The lattice Boltzmann
  equation for fluid dynamics and beyond}}},\ \bibinfo {edition} {1st}\ ed.\
  (\bibinfo  {publisher} {Clarendon, Oxford},\ \bibinfo {year}
  {2001})\BibitemShut {NoStop}%
\bibitem [{\citenamefont {Succi}(2015)}]{succi-2015}%
  \BibitemOpen
  \bibfield  {author} {\bibinfo {author} {\bibfnamefont {S.}~\bibnamefont
  {Succi}},\ }\bibfield  {title} {\enquote {\bibinfo {title} {Lattice boltzmann
  2038},}\ }\href@noop {} {\bibfield  {journal} {\bibinfo  {journal} {EPL
  (Europhysics Letters)}\ }\textbf {\bibinfo {volume} {109}},\ \bibinfo {pages}
  {50001} (\bibinfo {year} {2015})}\BibitemShut {NoStop}%
\bibitem [{\citenamefont {Benzi}, \citenamefont {Succi},\ and\ \citenamefont
  {Vergassola}(1992)}]{benzi-1992}%
  \BibitemOpen
  \bibfield  {author} {\bibinfo {author} {\bibfnamefont {R.}~\bibnamefont
  {Benzi}}, \bibinfo {author} {\bibfnamefont {S.}~\bibnamefont {Succi}}, \ and\
  \bibinfo {author} {\bibfnamefont {M.}~\bibnamefont {Vergassola}},\ }\bibfield
   {title} {\enquote {\bibinfo {title} {{The lattice Boltzmann equation: theory
  and applications}},}\ }\href@noop {} {\bibfield  {journal} {\bibinfo
  {journal} {Physics Reports}\ }\textbf {\bibinfo {volume} {222}},\ \bibinfo
  {pages} {145--197} (\bibinfo {year} {1992})}\BibitemShut {NoStop}%
\bibitem [{\citenamefont {Gonnella}\ and\ \citenamefont
  {Yeomans}(2009)}]{gonnellayeomans-2009}%
  \BibitemOpen
  \bibfield  {author} {\bibinfo {author} {\bibfnamefont {G.}~\bibnamefont
  {Gonnella}}\ and\ \bibinfo {author} {\bibfnamefont {J.~M.}\ \bibnamefont
  {Yeomans}},\ }\bibfield  {title} {\enquote {\bibinfo {title} {{Using the
  lattice Boltzmann algorithm to explore phase ordering in fluids.}}}\ }in\
  \href@noop {} {\emph {\bibinfo {booktitle} {Kinetics of Phase
  Transitions}}},\ \bibinfo {editor} {edited by\ \bibinfo {editor}
  {\bibfnamefont {S.}~\bibnamefont {Puri}}}\ (\bibinfo  {publisher} {CRC, Boca
  Raton},\ \bibinfo {year} {2009})\ pp.\ \bibinfo {pages} {89--166}\BibitemShut
  {NoStop}%
\bibitem [{\citenamefont {D\"{u}nweg}\ and\ \citenamefont
  {Ladd}(2009)}]{duenweg-2009}%
  \BibitemOpen
  \bibfield  {author} {\bibinfo {author} {\bibfnamefont {B.}~\bibnamefont
  {D\"{u}nweg}}\ and\ \bibinfo {author} {\bibfnamefont {A.}~\bibnamefont
  {Ladd}},\ }\bibfield  {title} {\enquote {\bibinfo {title} {Lattice boltzmann
  simulations of soft matter systems},}\ }in\ \href@noop {} {\emph {\bibinfo
  {booktitle} {Advanced Computer Simulation Approaches for Soft Matter Sciences
  III}}},\ \bibinfo {series} {Advances in Polymer Science}, Vol.\ \bibinfo
  {volume} {221},\ \bibinfo {editor} {edited by\ \bibinfo {editor}
  {\bibfnamefont {C.}~\bibnamefont {Holm}}\ and\ \bibinfo {editor}
  {\bibfnamefont {K.}~\bibnamefont {Kremer}}}\ (\bibinfo  {publisher} {Springer
  Berlin / Heidelberg},\ \bibinfo {year} {2009})\ pp.\ \bibinfo {pages}
  {89--166}\BibitemShut {NoStop}%
\bibitem [{\citenamefont {Sofonea}\ \emph {et~al.}(2004)\citenamefont
  {Sofonea}, \citenamefont {Lamura}, \citenamefont {Gonnella},\ and\
  \citenamefont {Cristea}}]{sofonea-2004}%
  \BibitemOpen
  \bibfield  {author} {\bibinfo {author} {\bibfnamefont {V.}~\bibnamefont
  {Sofonea}}, \bibinfo {author} {\bibfnamefont {A.}~\bibnamefont {Lamura}},
  \bibinfo {author} {\bibfnamefont {G.}~\bibnamefont {Gonnella}}, \ and\
  \bibinfo {author} {\bibfnamefont {A.}~\bibnamefont {Cristea}},\ }\bibfield
  {title} {\enquote {\bibinfo {title} {{Finite-difference Lattice Boltzmann
  Model with Flux Limiters for Liquid-vapor Systems}},}\ }\href@noop {}
  {\bibfield  {journal} {\bibinfo  {journal} {Phys. Rev. E}\ }\textbf {\bibinfo
  {volume} {70}},\ \bibinfo {pages} {046702} (\bibinfo {year}
  {2004})}\BibitemShut {NoStop}%
\bibitem [{\citenamefont {Cristea}\ \emph {et~al.}(2006)\citenamefont
  {Cristea}, \citenamefont {Gonnella}, \citenamefont {Lamura},\ and\
  \citenamefont {Sofonea}}]{cristea-2006}%
  \BibitemOpen
  \bibfield  {author} {\bibinfo {author} {\bibfnamefont {A.}~\bibnamefont
  {Cristea}}, \bibinfo {author} {\bibfnamefont {G.}~\bibnamefont {Gonnella}},
  \bibinfo {author} {\bibfnamefont {A.}~\bibnamefont {Lamura}}, \ and\ \bibinfo
  {author} {\bibfnamefont {V.}~\bibnamefont {Sofonea}},\ }\bibfield  {title}
  {\enquote {\bibinfo {title} {{Finite-difference Lattice Boltzmann Model for
  Liquid-vapor Systems}},}\ }\href@noop {} {\bibfield  {journal} {\bibinfo
  {journal} {Math. Comput. Simul.}\ }\textbf {\bibinfo {volume} {72}},\
  \bibinfo {pages} {113} (\bibinfo {year} {2006})}\BibitemShut {NoStop}%
\bibitem [{\citenamefont {Gonnella}, \citenamefont {Lamura},\ and\
  \citenamefont {Sofonea}(2007)}]{gonnella-2007}%
  \BibitemOpen
  \bibfield  {author} {\bibinfo {author} {\bibfnamefont {G.}~\bibnamefont
  {Gonnella}}, \bibinfo {author} {\bibfnamefont {A.}~\bibnamefont {Lamura}}, \
  and\ \bibinfo {author} {\bibfnamefont {V.}~\bibnamefont {Sofonea}},\
  }\bibfield  {title} {\enquote {\bibinfo {title} {{Lattice Boltzmann
  simulation of thermal nonideal fluids}},}\ }\href@noop {} {\bibfield
  {journal} {\bibinfo  {journal} {Phys. Rev. E}\ }\textbf {\bibinfo {volume}
  {76}},\ \bibinfo {pages} {036703} (\bibinfo {year} {2007})}\BibitemShut
  {NoStop}%
\bibitem [{\citenamefont {Gonnella}, \citenamefont {Lamura},\ and\
  \citenamefont {Sofonea}(2009)}]{gonnella-2009}%
  \BibitemOpen
  \bibfield  {author} {\bibinfo {author} {\bibfnamefont {G.}~\bibnamefont
  {Gonnella}}, \bibinfo {author} {\bibfnamefont {A.}~\bibnamefont {Lamura}}, \
  and\ \bibinfo {author} {\bibfnamefont {V.}~\bibnamefont {Sofonea}},\
  }\bibfield  {title} {\enquote {\bibinfo {title} {{A lattice Boltzmann method
  for thermal nonideal fluids}},}\ }\href@noop {} {\bibfield  {journal}
  {\bibinfo  {journal} {Eur. Phys. J.-Spec. Top.}\ }\textbf {\bibinfo {volume}
  {171}},\ \bibinfo {pages} {181} (\bibinfo {year} {2009})}\BibitemShut
  {NoStop}%
\bibitem [{\citenamefont {Cristea}\ \emph {et~al.}(2010)\citenamefont
  {Cristea}, \citenamefont {Gonnella}, \citenamefont {Lamura},\ and\
  \citenamefont {Sofonea}}]{cristea-2010}%
  \BibitemOpen
  \bibfield  {author} {\bibinfo {author} {\bibfnamefont {A.}~\bibnamefont
  {Cristea}}, \bibinfo {author} {\bibfnamefont {G.}~\bibnamefont {Gonnella}},
  \bibinfo {author} {\bibfnamefont {A.}~\bibnamefont {Lamura}}, \ and\ \bibinfo
  {author} {\bibfnamefont {V.}~\bibnamefont {Sofonea}},\ }\bibfield  {title}
  {\enquote {\bibinfo {title} {{A lattice Boltzmann study of phase separation
  in liquid-vapor systems with gravity}},}\ }\href@noop {} {\bibfield
  {journal} {\bibinfo  {journal} {Commun. Comput. Phys.}\ }\textbf {\bibinfo
  {volume} {7}},\ \bibinfo {pages} {350} (\bibinfo {year} {2010})}\BibitemShut
  {NoStop}%
\bibitem [{\citenamefont {Falcucci}\ \emph
  {et~al.}(2013{\natexlab{a}})\citenamefont {Falcucci}, \citenamefont
  {Jannelli}, \citenamefont {Ubertini},\ and\ \citenamefont
  {Succi}}]{falcucci-2013}%
  \BibitemOpen
  \bibfield  {author} {\bibinfo {author} {\bibfnamefont {G.}~\bibnamefont
  {Falcucci}}, \bibinfo {author} {\bibfnamefont {E.}~\bibnamefont {Jannelli}},
  \bibinfo {author} {\bibfnamefont {S.}~\bibnamefont {Ubertini}}, \ and\
  \bibinfo {author} {\bibfnamefont {S.}~\bibnamefont {Succi}},\ }\bibfield
  {title} {\enquote {\bibinfo {title} {{Direct numerical evidence of
  stress-induced cavitation}},}\ }\href@noop {} {\bibfield  {journal} {\bibinfo
   {journal} {Journal of Fluid Mechanics}\ }\textbf {\bibinfo {volume} {728}},\
  \bibinfo {pages} {362} (\bibinfo {year} {2013}{\natexlab{a}})}\BibitemShut
  {NoStop}%
\bibitem [{\citenamefont {Falcucci}\ \emph
  {et~al.}(2013{\natexlab{b}})\citenamefont {Falcucci}, \citenamefont
  {Ubertini}, \citenamefont {Bella},\ and\ \citenamefont
  {Succi}}]{falcucci-2013a}%
  \BibitemOpen
  \bibfield  {author} {\bibinfo {author} {\bibfnamefont {G.}~\bibnamefont
  {Falcucci}}, \bibinfo {author} {\bibfnamefont {S.}~\bibnamefont {Ubertini}},
  \bibinfo {author} {\bibfnamefont {G.}~\bibnamefont {Bella}}, \ and\ \bibinfo
  {author} {\bibfnamefont {S.}~\bibnamefont {Succi}},\ }\bibfield  {title}
  {\enquote {\bibinfo {title} {{Lattice Boltzmann simulation of cavitating
  flows}},}\ }\href@noop {} {\bibfield  {journal} {\bibinfo  {journal} {Commun.
  Comput. Phys.}\ }\textbf {\bibinfo {volume} {13}},\ \bibinfo {pages}
  {685--695} (\bibinfo {year} {2013}{\natexlab{b}})}\BibitemShut {NoStop}%
\bibitem [{\citenamefont {Sukop}\ and\ \citenamefont {Or}(2005)}]{sukop-2005}%
  \BibitemOpen
  \bibfield  {author} {\bibinfo {author} {\bibfnamefont {M.~C.}\ \bibnamefont
  {Sukop}}\ and\ \bibinfo {author} {\bibfnamefont {D.}~\bibnamefont {Or}},\
  }\bibfield  {title} {\enquote {\bibinfo {title} {Lattice boltzmann method for
  homogeneous and heterogeneous cavitation},}\ }\href {\doibase
  10.1103/PhysRevE.71.046703} {\bibfield  {journal} {\bibinfo  {journal} {Phys.
  Rev. E}\ }\textbf {\bibinfo {volume} {71}},\ \bibinfo {pages} {046703}
  (\bibinfo {year} {2005})}\BibitemShut {NoStop}%
\bibitem [{\citenamefont {Chen}, \citenamefont {Zhong},\ and\ \citenamefont
  {Yuan}(2011)}]{xpchen-2011}%
  \BibitemOpen
  \bibfield  {author} {\bibinfo {author} {\bibfnamefont {X.-P.}\ \bibnamefont
  {Chen}}, \bibinfo {author} {\bibfnamefont {C.-W.}\ \bibnamefont {Zhong}}, \
  and\ \bibinfo {author} {\bibfnamefont {X.-L.}\ \bibnamefont {Yuan}},\
  }\bibfield  {title} {\enquote {\bibinfo {title} {Lattice boltzmann simulation
  of cavitating bubble growth with large density ratio},}\ }\href@noop {}
  {\bibfield  {journal} {\bibinfo  {journal} {Computers and Mathematics with
  Applications}\ }\textbf {\bibinfo {volume} {61}},\ \bibinfo {pages} {3577 --
  3584} (\bibinfo {year} {2011})}\BibitemShut {NoStop}%
\bibitem [{\citenamefont {Zhong}, \citenamefont {Zhong},\ and\ \citenamefont
  {Bai}(2012)}]{zhong-2012}%
  \BibitemOpen
  \bibfield  {author} {\bibinfo {author} {\bibfnamefont {M.}~\bibnamefont
  {Zhong}}, \bibinfo {author} {\bibfnamefont {C.}~\bibnamefont {Zhong}}, \ and\
  \bibinfo {author} {\bibfnamefont {C.}~\bibnamefont {Bai}},\ }\bibfield
  {title} {\enquote {\bibinfo {title} {A high-order discrete scheme of lattice
  boltzmann method for cavitation simulations.}}\ }\href@noop {} {\bibfield
  {journal} {\bibinfo  {journal} {Adv. Comput. Sci. Appls}\ }\textbf {\bibinfo
  {volume} {1}},\ \bibinfo {pages} {73--77} (\bibinfo {year}
  {2012})}\BibitemShut {NoStop}%
\bibitem [{\citenamefont {Falcucci}\ \emph {et~al.}(2007)\citenamefont
  {Falcucci}, \citenamefont {Bella}, \citenamefont {Chiatti}, \citenamefont
  {Chibbaro}, \citenamefont {Sbragaglia},\ and\ \citenamefont
  {Succi}}]{falcucci-2007}%
  \BibitemOpen
  \bibfield  {author} {\bibinfo {author} {\bibfnamefont {G.}~\bibnamefont
  {Falcucci}}, \bibinfo {author} {\bibfnamefont {G.}~\bibnamefont {Bella}},
  \bibinfo {author} {\bibfnamefont {G.}~\bibnamefont {Chiatti}}, \bibinfo
  {author} {\bibfnamefont {S.}~\bibnamefont {Chibbaro}}, \bibinfo {author}
  {\bibfnamefont {M.}~\bibnamefont {Sbragaglia}}, \ and\ \bibinfo {author}
  {\bibfnamefont {S.}~\bibnamefont {Succi}},\ }\bibfield  {title} {\enquote
  {\bibinfo {title} {{Lattice Boltzmann models with mid-range interactions}},}\
  }\href@noop {} {\bibfield  {journal} {\bibinfo  {journal} {Commun. Comput.
  Phys.}\ }\textbf {\bibinfo {volume} {2}},\ \bibinfo {pages} {1071--1084}
  (\bibinfo {year} {2007})}\BibitemShut {NoStop}%
\bibitem [{\citenamefont {Shan}\ and\ \citenamefont {Chen}(1993)}]{shan-1993}%
  \BibitemOpen
  \bibfield  {author} {\bibinfo {author} {\bibfnamefont {X.}~\bibnamefont
  {Shan}}\ and\ \bibinfo {author} {\bibfnamefont {H.}~\bibnamefont {Chen}},\
  }\bibfield  {title} {\enquote {\bibinfo {title} {{Lattice Boltzmann model for
  simulating flows with multiple phases and components}},}\ }\href@noop {}
  {\bibfield  {journal} {\bibinfo  {journal} {Phys. Rev. E}\ }\textbf {\bibinfo
  {volume} {47}},\ \bibinfo {pages} {1815--1819} (\bibinfo {year}
  {1993})}\BibitemShut {NoStop}%
\bibitem [{\citenamefont {Coclite}, \citenamefont {Gonnella},\ and\
  \citenamefont {Lamura}(2014)}]{coclite-2014}%
  \BibitemOpen
  \bibfield  {author} {\bibinfo {author} {\bibfnamefont {A.}~\bibnamefont
  {Coclite}}, \bibinfo {author} {\bibfnamefont {G.}~\bibnamefont {Gonnella}}, \
  and\ \bibinfo {author} {\bibfnamefont {A.}~\bibnamefont {Lamura}},\
  }\bibfield  {title} {\enquote {\bibinfo {title} {Pattern formation in
  liquid-vapor systems under periodic potential and shear},}\ }\href@noop {}
  {\bibfield  {journal} {\bibinfo  {journal} {Physical Review E}\ }\textbf
  {\bibinfo {volume} {89}},\ \bibinfo {pages} {063303} (\bibinfo {year}
  {2014})}\BibitemShut {NoStop}%
\bibitem [{\citenamefont {Shan}, \citenamefont {Yuan},\ and\ \citenamefont
  {Chen}(2006)}]{shan-2006}%
  \BibitemOpen
  \bibfield  {author} {\bibinfo {author} {\bibfnamefont {X.}~\bibnamefont
  {Shan}}, \bibinfo {author} {\bibfnamefont {X.}~\bibnamefont {Yuan}}, \ and\
  \bibinfo {author} {\bibfnamefont {H.}~\bibnamefont {Chen}},\ }\bibfield
  {title} {\enquote {\bibinfo {title} {{Kinetic theory representation of
  hydrodynamics: a way beyond the Navier Stokes equation}},}\ }\href@noop {}
  {\bibfield  {journal} {\bibinfo  {journal} {J. Fluid Mech.}\ }\textbf
  {\bibinfo {volume} {550}},\ \bibinfo {pages} {413--441} (\bibinfo {year}
  {2006})}\BibitemShut {NoStop}%
\bibitem [{\citenamefont {Guo}, \citenamefont {Zheng},\ and\ \citenamefont
  {Shi}(2002)}]{guo-2002}%
  \BibitemOpen
  \bibfield  {author} {\bibinfo {author} {\bibfnamefont {Z.}~\bibnamefont
  {Guo}}, \bibinfo {author} {\bibfnamefont {C.}~\bibnamefont {Zheng}}, \ and\
  \bibinfo {author} {\bibfnamefont {B.}~\bibnamefont {Shi}},\ }\bibfield
  {title} {\enquote {\bibinfo {title} {Discrete lattice effects on the forcing
  term in the lattice boltzmann method},}\ }\href@noop {} {\bibfield  {journal}
  {\bibinfo  {journal} {Physical Review E}\ }\textbf {\bibinfo {volume} {65}},\
  \bibinfo {pages} {46308} (\bibinfo {year} {2002})}\BibitemShut {NoStop}%
\bibitem [{\citenamefont {Tiribocchi}\ \emph {et~al.}(2009)\citenamefont
  {Tiribocchi}, \citenamefont {Stella}, \citenamefont {Gonnella},\ and\
  \citenamefont {Lamura}}]{tiribocchi-2009}%
  \BibitemOpen
  \bibfield  {author} {\bibinfo {author} {\bibfnamefont {A.}~\bibnamefont
  {Tiribocchi}}, \bibinfo {author} {\bibfnamefont {N.}~\bibnamefont {Stella}},
  \bibinfo {author} {\bibfnamefont {G.}~\bibnamefont {Gonnella}}, \ and\
  \bibinfo {author} {\bibfnamefont {A.}~\bibnamefont {Lamura}},\ }\bibfield
  {title} {\enquote {\bibinfo {title} {{Hybrid lattice Boltzmann model for
  binary fluid mixtures}},}\ }\href@noop {} {\bibfield  {journal} {\bibinfo
  {journal} {Phys. Rev. E}\ }\textbf {\bibinfo {volume} {80}},\ \bibinfo
  {pages} {026701} (\bibinfo {year} {2009})}\BibitemShut {NoStop}%
\bibitem [{\citenamefont {Gonnella}\ \emph {et~al.}(2010)\citenamefont
  {Gonnella}, \citenamefont {Lamura}, \citenamefont {Piscitelli},\ and\
  \citenamefont {Tiribocchi}}]{gonnella-2010}%
  \BibitemOpen
  \bibfield  {author} {\bibinfo {author} {\bibfnamefont {G.}~\bibnamefont
  {Gonnella}}, \bibinfo {author} {\bibfnamefont {A.}~\bibnamefont {Lamura}},
  \bibinfo {author} {\bibfnamefont {A.}~\bibnamefont {Piscitelli}}, \ and\
  \bibinfo {author} {\bibfnamefont {A.}~\bibnamefont {Tiribocchi}},\ }\bibfield
   {title} {\enquote {\bibinfo {title} {Phase separation of binary fluids with
  dynamic temperature},}\ }\href@noop {} {\bibfield  {journal} {\bibinfo
  {journal} {Phys. Rev. E}\ }\textbf {\bibinfo {volume} {82}},\ \bibinfo
  {pages} {046302} (\bibinfo {year} {2010})}\BibitemShut {NoStop}%
\bibitem [{\citenamefont {Gonnella}, \citenamefont {Lamura},\ and\
  \citenamefont {Tiribocchi}(2011)}]{gonnella-2011}%
  \BibitemOpen
  \bibfield  {author} {\bibinfo {author} {\bibfnamefont {G.}~\bibnamefont
  {Gonnella}}, \bibinfo {author} {\bibfnamefont {A.}~\bibnamefont {Lamura}}, \
  and\ \bibinfo {author} {\bibfnamefont {A.}~\bibnamefont {Tiribocchi}},\
  }\bibfield  {title} {\enquote {\bibinfo {title} {Thermal and hydrodynamic
  effects in the ordering of lamellar fluids},}\ }\href@noop {} {\bibfield
  {journal} {\bibinfo  {journal} {Philos. Trans. R. Soc. A}\ }\textbf {\bibinfo
  {volume} {369}},\ \bibinfo {pages} {2592} (\bibinfo {year}
  {2011})}\BibitemShut {NoStop}%
\bibitem [{\citenamefont {Rowlinson}\ and\ \citenamefont
  {Widom}(1982)}]{rowlinson-1982}%
  \BibitemOpen
  \bibfield  {author} {\bibinfo {author} {\bibfnamefont {J.~S.}\ \bibnamefont
  {Rowlinson}}\ and\ \bibinfo {author} {\bibfnamefont {B.}~\bibnamefont
  {Widom}},\ }\href@noop {} {\emph {\bibinfo {title} {Molecular Theory of
  Capillarity}}}\ (\bibinfo  {publisher} {Clarendon, Oxford},\ \bibinfo {year}
  {1982})\BibitemShut {NoStop}%
\bibitem [{\citenamefont {Shan}(2006)}]{shan-2006b}%
  \BibitemOpen
  \bibfield  {author} {\bibinfo {author} {\bibfnamefont {X.}~\bibnamefont
  {Shan}},\ }\bibfield  {title} {\enquote {\bibinfo {title} {{Analysis and
  reduction of the spurious current in a class of multiphase lattice Boltzmann
  models}},}\ }\href@noop {} {\bibfield  {journal} {\bibinfo  {journal} {Phys.
  Rev. E}\ }\textbf {\bibinfo {volume} {73}},\ \bibinfo {pages} {047701}
  (\bibinfo {year} {2006})}\BibitemShut {NoStop}%
\bibitem [{\citenamefont {Sbragaglia}\ \emph {et~al.}(2007)\citenamefont
  {Sbragaglia}, \citenamefont {Benzi}, \citenamefont {Biferale}, \citenamefont
  {Succi}, \citenamefont {Sugiyama},\ and\ \citenamefont
  {Toschi}}]{sbragaglia-2007}%
  \BibitemOpen
  \bibfield  {author} {\bibinfo {author} {\bibfnamefont {M.}~\bibnamefont
  {Sbragaglia}}, \bibinfo {author} {\bibfnamefont {R.}~\bibnamefont {Benzi}},
  \bibinfo {author} {\bibfnamefont {L.}~\bibnamefont {Biferale}}, \bibinfo
  {author} {\bibfnamefont {S.}~\bibnamefont {Succi}}, \bibinfo {author}
  {\bibfnamefont {K.}~\bibnamefont {Sugiyama}}, \ and\ \bibinfo {author}
  {\bibfnamefont {F.}~\bibnamefont {Toschi}},\ }\bibfield  {title} {\enquote
  {\bibinfo {title} {{Generalized lattice Boltzmann method with multirange
  pseudopotential}},}\ }\href@noop {} {\bibfield  {journal} {\bibinfo
  {journal} {Phys. Rev. E}\ }\textbf {\bibinfo {volume} {75}},\ \bibinfo
  {pages} {026702} (\bibinfo {year} {2007})}\BibitemShut {NoStop}%
\bibitem [{\citenamefont {Nurick}(1976)}]{nurick-1976}%
  \BibitemOpen
  \bibfield  {author} {\bibinfo {author} {\bibfnamefont {W.~H.}\ \bibnamefont
  {Nurick}},\ }\bibfield  {title} {\enquote {\bibinfo {title} {Orifice
  cavitation and its effect on spray mixing},}\ }\href@noop {} {\bibfield
  {journal} {\bibinfo  {journal} {J Fluid Eng}\ }\textbf {\bibinfo {volume}
  {98}},\ \bibinfo {pages} {681--687} (\bibinfo {year} {1976})}\BibitemShut
  {NoStop}%
\bibitem [{\citenamefont {Lamura}\ and\ \citenamefont
  {Gonnella}(2001)}]{lamura-2001}%
  \BibitemOpen
  \bibfield  {author} {\bibinfo {author} {\bibfnamefont {A.}~\bibnamefont
  {Lamura}}\ and\ \bibinfo {author} {\bibfnamefont {G.}~\bibnamefont
  {Gonnella}},\ }\bibfield  {title} {\enquote {\bibinfo {title} {Lattice
  boltzmann simulations of segregating binary fluid mixtures in shear flow},}\
  }\href@noop {} {\bibfield  {journal} {\bibinfo  {journal} {Physica A}\
  }\textbf {\bibinfo {volume} {294}},\ \bibinfo {pages} {295} (\bibinfo {year}
  {2001})}\BibitemShut {NoStop}%
\bibitem [{\citenamefont {Tiribocchi}\ \emph {et~al.}(2011)\citenamefont
  {Tiribocchi}, \citenamefont {Piscitelli}, \citenamefont {Gonnella},\ and\
  \citenamefont {Lamura}}]{tiribocchi-2011}%
  \BibitemOpen
  \bibfield  {author} {\bibinfo {author} {\bibfnamefont {A.}~\bibnamefont
  {Tiribocchi}}, \bibinfo {author} {\bibfnamefont {A.}~\bibnamefont
  {Piscitelli}}, \bibinfo {author} {\bibfnamefont {G.}~\bibnamefont
  {Gonnella}}, \ and\ \bibinfo {author} {\bibfnamefont {A.}~\bibnamefont
  {Lamura}},\ }\bibfield  {title} {\enquote {\bibinfo {title} {Pattern study of
  thermal phase separation for binary fluid mixtures},}\ }\href@noop {}
  {\bibfield  {journal} {\bibinfo  {journal} {Int. J. Numer. Methods Heat Fluid
  Flow}\ }\textbf {\bibinfo {volume} {21}},\ \bibinfo {pages} {572} (\bibinfo
  {year} {2011})}\BibitemShut {NoStop}%
\bibitem [{\citenamefont {Wagner}\ and\ \citenamefont
  {Pooley}(2007)}]{wagner-2007}%
  \BibitemOpen
  \bibfield  {author} {\bibinfo {author} {\bibfnamefont {A.~J.}\ \bibnamefont
  {Wagner}}\ and\ \bibinfo {author} {\bibfnamefont {C.~M.}\ \bibnamefont
  {Pooley}},\ }\bibfield  {title} {\enquote {\bibinfo {title} {Interface width
  and bulk stability: Requirements for the simulation of deeply quenched
  liquid-gas systems},}\ }\href@noop {} {\bibfield  {journal} {\bibinfo
  {journal} {Phys. Rev. E}\ }\textbf {\bibinfo {volume} {76}},\ \bibinfo
  {pages} {045702(R)} (\bibinfo {year} {2007})}\BibitemShut {NoStop}%
\bibitem [{\citenamefont {Gonnella}\ and\ \citenamefont
  {Pellicoro}(2000)}]{gonnella-2000}%
  \BibitemOpen
  \bibfield  {author} {\bibinfo {author} {\bibfnamefont {G.}~\bibnamefont
  {Gonnella}}\ and\ \bibinfo {author} {\bibfnamefont {M.}~\bibnamefont
  {Pellicoro}},\ }\bibfield  {title} {\enquote {\bibinfo {title} {Critical
  temperatures in driven binary mixtures with conserved and non-conserved
  dynamics},}\ }\href@noop {} {\bibfield  {journal} {\bibinfo  {journal}
  {Journal of Physics. A}\ }\textbf {\bibinfo {volume} {33}},\ \bibinfo {pages}
  {7043--7051} (\bibinfo {year} {2000})}\BibitemShut {NoStop}%
\bibitem [{\citenamefont {Corberi}, \citenamefont {Gonnella},\ and\
  \citenamefont {Lamura}(1998)}]{gonnella-1998}%
  \BibitemOpen
  \bibfield  {author} {\bibinfo {author} {\bibfnamefont {F.}~\bibnamefont
  {Corberi}}, \bibinfo {author} {\bibfnamefont {G.}~\bibnamefont {Gonnella}}, \
  and\ \bibinfo {author} {\bibfnamefont {A.}~\bibnamefont {Lamura}},\
  }\bibfield  {title} {\enquote {\bibinfo {title} {Spinodal decomposition of
  binary mixtures in uniform shear flow},}\ }\href@noop {} {\bibfield
  {journal} {\bibinfo  {journal} {Phys. Rev. Lett.}\ }\textbf {\bibinfo
  {volume} {81}},\ \bibinfo {pages} {3852} (\bibinfo {year}
  {1998})}\BibitemShut {NoStop}%
\bibitem [{Note1()}]{Note1}%
  \BibitemOpen
  \bibinfo {note} {The final increase in the stress seen before cavitation is
  due to the contribution of the van der Waals pressure that, following the van
  der Waals isotherm, increases when the fluid density decrease below the
  liquid spinodal value.}\BibitemShut {Stop}%
\bibitem [{\citenamefont {Patanastasiou}, \citenamefont {Georgiou},\ and\
  \citenamefont {Alexandrou}(1999)}]{papanastasiou-1999}%
  \BibitemOpen
  \bibfield  {author} {\bibinfo {author} {\bibfnamefont {T.}~\bibnamefont
  {Patanastasiou}}, \bibinfo {author} {\bibfnamefont {G.}~\bibnamefont
  {Georgiou}}, \ and\ \bibinfo {author} {\bibfnamefont {A.}~\bibnamefont
  {Alexandrou}},\ }\href@noop {} {\emph {\bibinfo {title} {Viscous Fluid
  Flow}}}\ (\bibinfo  {publisher} {CRC, Boca Raton},\ \bibinfo {year}
  {1999})\BibitemShut {NoStop}%
\bibitem [{\citenamefont {Schlichting}(1979)}]{schlichting-1979}%
  \BibitemOpen
  \bibfield  {author} {\bibinfo {author} {\bibfnamefont {H.}~\bibnamefont
  {Schlichting}},\ }\href@noop {} {\emph {\bibinfo {title} {Boundary layer
  theory}}},\ \bibinfo {edition} {7th}\ ed.\ (\bibinfo  {publisher}
  {McGraw-Hill Book Company},\ \bibinfo {address} {New York},\ \bibinfo {year}
  {1979})\BibitemShut {NoStop}%
\bibitem [{\citenamefont {Wagner}\ and\ \citenamefont
  {Yeomans}(1999)}]{wagner-1999}%
  \BibitemOpen
  \bibfield  {author} {\bibinfo {author} {\bibfnamefont {A.~J.}\ \bibnamefont
  {Wagner}}\ and\ \bibinfo {author} {\bibfnamefont {J.~M.}\ \bibnamefont
  {Yeomans}},\ }\bibfield  {title} {\enquote {\bibinfo {title} {Phase
  separation under shear in two-dimensional binary fluids},}\ }\href@noop {}
  {\bibfield  {journal} {\bibinfo  {journal} {Phys. Rev. E}\ }\textbf {\bibinfo
  {volume} {59}},\ \bibinfo {pages} {4366--4373} (\bibinfo {year}
  {1999})}\BibitemShut {NoStop}%
\bibitem [{\citenamefont {Corberi}, \citenamefont {Gonnella},\ and\
  \citenamefont {Lamura}(1999)}]{gonnella-1999}%
  \BibitemOpen
  \bibfield  {author} {\bibinfo {author} {\bibfnamefont {F.}~\bibnamefont
  {Corberi}}, \bibinfo {author} {\bibfnamefont {G.}~\bibnamefont {Gonnella}}, \
  and\ \bibinfo {author} {\bibfnamefont {A.}~\bibnamefont {Lamura}},\
  }\bibfield  {title} {\enquote {\bibinfo {title} {Two-scale competition in
  phase separation with shear},}\ }\href@noop {} {\bibfield  {journal}
  {\bibinfo  {journal} {Phys. Rev. Lett.}\ }\textbf {\bibinfo {volume} {83}},\
  \bibinfo {pages} {4057} (\bibinfo {year} {1999})}\BibitemShut {NoStop}%
\bibitem [{\citenamefont {Wagner}\ and\ \citenamefont
  {Li}(2006)}]{wagner-2006}%
  \BibitemOpen
  \bibfield  {author} {\bibinfo {author} {\bibfnamefont {A.~J.}\ \bibnamefont
  {Wagner}}\ and\ \bibinfo {author} {\bibfnamefont {Q.}~\bibnamefont {Li}},\
  }\bibfield  {title} {\enquote {\bibinfo {title} {Investigation of galilean
  invariance of multi-phase lattice boltzmann methods},}\ }\href@noop {}
  {\bibfield  {journal} {\bibinfo  {journal} {Physica A}\ }\textbf {\bibinfo
  {volume} {362}},\ \bibinfo {pages} {105--110} (\bibinfo {year}
  {2006})}\BibitemShut {NoStop}%
\end{thebibliography}

%

\end{document}